\documentclass[pre,reprint,showpacs]{revtex4-1}
\usepackage{amssymb,amsmath}
\usepackage{natbib}
\usepackage{graphicx}
\usepackage{color,soul}
\usepackage{bm}

\newcommand\bra[1]{{\langle{#1}|}}
\newcommand\ket[1]{{|{#1}\rangle}}

	 \renewcommand{\vec}[1]{\bm{#1}}

  \def \cl {\text{cl}}
  \def \ing {\text{int}}
  
\begin{document}

\title{ Quasi-integrable systems  are  slow to thermalize but may  be good scramblers}

\author{Tomer Goldfriend}
\email{tomergf@gmail.com}
\affiliation{Laboratoire  de  Physique  Statistique, D\'epartement de physique de l\'ENS, \'Ecole  Normale  Sup\'erieure,
PSL  Research  University;  Universit\'e  Paris  Diderot, Sorbonne  Paris-Cit\'e;  Sorbonne
Universit\'es,  UPMC  Univ.  Paris  06,  CNRS;  24  rue  Lhomond,  75005  Paris,  France}

\author{Jorge Kurchan} 
\affiliation{Laboratoire  de  Physique  Statistique, D\'epartement de physique de l\'ENS, \'Ecole  Normale  Sup\'erieure,
PSL  Research  University;  Universit\'e  Paris  Diderot, Sorbonne  Paris-Cit\'e;  Sorbonne
Universit\'es,  UPMC  Univ.  Paris  06,  CNRS;  24  rue  Lhomond,  75005  Paris,  France}

\date{\today}

\begin{abstract}

{ Classical quasi-integrable systems are known to have Lyapunov times much shorter than their ergodicity time --  the most clear example being the Solar System --
 but the situation for their quantum counterparts is less
well understood. As a first example, we examine the quantum Lyapunov exponent, defined by the evolution of the 4-point out-of-time-order correlator (OTOC), of integrable systems which are weakly perturbed by an external noise, a setting that has proven to be illuminating in the classical case. In analogy to the tangent space in classical systems, we derive a linear superoperator equation which dictates the OTOC dynamics. We find that {\em i)} in the semi-classical limit the quantum Lyapunov exponent is given by the classical one: it  scales as $\epsilon^{1/3}$, with $\epsilon$ being the variance of the random drive, leading to short Lyapunov times compared to the diffusion time (which is $\sim \epsilon^{-1}$). {\em ii)} in the highly quantal regime the Lyapunov instability is suppressed by quantum  fluctuations, and {\em iii)} for sufficiently small perturbations the $\epsilon^{1/3}$ dependence is also suppressed -- another purely quantum effect which we explain.
These essential features of the problem are already present in a  rotor that is  kicked {\em weakly} but {\em randomly}. Concerning  quantum limits on chaos, we find that quasi-integrable systems are relatively good scramblers in the sense that the ratio between the Lyapunov exponent and $kT/\hbar$ may stay finite at a low temperature~$T$. 
}
\end{abstract}


\maketitle

\section{Introduction}
\label{sec:intro}

The field of Quantum Chaos was born from the attempt to understand how the characteristics of classical chaotic systems appear in Quantum Mechanics~\cite{Casati&Chirikov}. By definition, Classical Chaos refers to high sensitivity to initial conditions. This is traditionally measured by the largest Lyapunov exponent that gives the exponential rate at which two initially close by trajectories separate in time. Given the lack of an equivalent measure in the linear and unitary quantum evolution, in 1984, Peres~\cite{Peres1984} suggested a Loschmidt echo protocol (fidelity) as an analog quantity to characterized Quantum Chaos. It was only about 15 years later, when new echo experiments and interest in quantum computation motivated theorists (original work in Ref.~\cite{Jalabert&Pastawski2001}) to show how the fidelity is connected to the classical Lyapunov exponent in the semi-classical limit. A slightly different measure that was already introduced in the late 60s~\cite{Larkin&Ovchinnikov1969}--- the Out-of-Time-Order-Correlator (OTOC)--- is in the focus of a recent revival in the field of Quantum Chaos. The OTOC is a 4-point correlation function referring to the square of the commutation relation between operators at time $t$ and time zero, $\langle[A(t),B_0]^2\rangle$, where the average is usually taken over thermal ensemble.

Semi-classical approximations~\cite{Larkin&Ovchinnikov1969,Maldacena_etal2016}, quantum information scrambling~\cite{Kitaev,Polchinski_etal2016,Maldacena&Stanford2016}, and a direct relation to the Loschmidt echo~\cite{Kurchan2018} connect between chaoticity and an exponential growth of the OTOC--- leading to the term Quantum Lyapunov exponent $\lambda_{Q}$, whenever $\langle[A(t),B_0]^2\rangle\sim e^{2\lambda_Q t}$.  The recent wave of  studies on the OTOC was initiated in the context of holographic theories for black holes, in the from  of a quantum bound on its growth~\cite{Maldacena_etal2016}. This theoretical interest is accompanied with new experimental abilities for controlling cold-atom systems~\cite{Bloch_etal2008,Eisert_etal2015} (and references therein). These set-ups concern many-body {\it isolated} systems, which might be integrable~\cite{Kinoshita_etal2004}, near-integrable~\cite{Langen_etal2016,Tang_etal2018}, or chaotic~\cite{D'Alessio_etal2016}. They allow one to study the behavior of the OTOC and its relevance to different processes in closed many-body quantum systems. Several experimental realizations to measure the OTOC were suggested~\cite{Bohrdt_etal2017,Swingle_etal2016,Yao_etal2016,Zhu_etal2016,Halpern2017,Wei_etal2018,Dag&Duan2019} and preformed~\cite{Li_etal2017,Garttner_etal2017,Landsman_etal2019}.

In Classical Mechanics, many-body near-integrable (or, quasi-integrable) systems exhibit strong chaotic behavior but  thermalize slowly,  the Lyapunov time characterizing the former is much smaller than the phase-space diffusion time associated with  the latter. A well-known example is the Solar System, which has a Lyapunov time of $\sim 5$ Myrs and stability time of $> 5$ Gyrs~\cite{Laskar2008}. The resulting relaxation process  involves slow dynamics from one ergodized torus to the another. The separation between  chaotic and ergodic time-scales can be understood  because the  Lyapunov instability is mostly tangent to the high-dimensional invariant tori, and hence helps little with thermalization. This was illustrated for the paradigmatic quasi-integrable system--- the Fermi-Pasta-Ulam-Tsingou chain--- where the route to equilibration passes through quasi-static states that live on invariant tori of the integrable Toda chain~\cite{Goldfriend&Kurchan2019}. 

The behavior of quantum quasi-integrable systems is akin to the classical one. A quantum protocol that has received extensive attention is to follow the thermalization of an isolated system  starting from some initial state: quasi-integrable systems quickly evolve to a long-lived {\em prethermalized state} determined by the (quantum) quasi-constants of motion,  followed by a slow relaxation to equilibrium~\cite{Langen_etal2016,Tang_etal2018}. 
 In general, `thermalization time' refers to the time it takes for a wavepacket to explore sequentially space with equilibrium probability, irrespective of its size. When  we refer to `scrambling' / Ehrenfest time, we mean the time  it takes for the packet to be spread over all accessible space at each time.  Clearly, the latter may be  infinite in a strictly classical situation, while  the former is typically finite, even classically. In this paper, we apply the same difference to `prethermalization' and `{\em prescrambling}', where the space in question is the torus of prethermalizad Hilbert space. 

For the classical problem, the essence of the dynamics of quasi-integrable systems may be understood  with an analytically  much simpler example: an integrable system which is weakly perturbed by an external noise~\cite{Lam&Kurchan2014} -- arguably, a stochastic drive can simulate the effect of the many-body integrability-breaking interactions. In particular, Ref.~\cite{Lam&Kurchan2014} showed that the randomly driven system develops chaos that is almost tangent to the invariant high-dimensional torus, with a Lyapunov time which is much smaller than the diffusion time. In addition, chaos appears for any magnitude of the noise (no regular islands in phase-space for any value of perturbation), thus the stochastic model can mimic the behavior of classical quasi-integrable systems beyond the Kolmogorov-Arnold-Moser (KAM) regime. It also does not contradict the KAM theorem, as an external random drive can be thought as coupling to an infinite set of oscillators with all frequencies~\cite{Zwanzig1973},
thus `resonating with everything'.

In the current Paper the ideas described in the last paragraph are extended to quantum systems. We study in detail chaos in quantum integrable systems which are weakly perturbed by (classical) noise. 
The initial conditions we have in mind will be  linear combinations of states $\sum_\alpha c_\alpha |\alpha \rangle$ having a set of quantum numbers 
$\{I_\alpha^1,...,I_\alpha^n\} \sim \{I_o^1,...,I_o^n\} $ that  correspond
to approximately equal values of all the constants of motion -- the quantum analogue of starting `near a torus'. Time evolution will
dephase these contributions, even before the breaking of integrability makes the amplitude norms change appreciably, i.e.
$ |c_\alpha| e^{i\psi_\alpha} \rightarrow  |c_\alpha| e^{i\psi_\alpha' }$  . The role
of `chaos on the torus' for classical systems is now played by `dephasing at constant $|c_\alpha|$' .
Remarkably, the noise term has a strong effect on dephasing, even before the quantum numbers have changed substantially.
An important timescale in the present paper is the scrambling-on-the-torus time--- the {\em prescrambling} time---   at which the $e^{i\psi_\alpha' }$ are essentially
random, but the $ |c_\alpha|$ have not yet diffused: this is the time it takes for a wavepacket to cover the whole torus.
For these timescales, we shall  focus on the evolution of the  OTOC, and  on the quantum bounds on chaos. 


The paper is organized as follows: First, we set the framework for the analysis by indicating notations in Sec.~\ref{sec:notation}; defining the quantum Lyapunov exponent together with the quantum tangent space in Sec.~\ref{sec:tangent}; and presenting our general stochastic one-dimensional quantum model for quasi-integrable systems in Sec.~\ref{sec:model}. Then, Sec.~\ref{sec:classical} summarizes the solution to the classical model~\cite{Lam&Kurchan2014}, and is followed by Sec.~\ref{sec:summ}, in which we outline our main results concerning the quantum model. In Sec.~\ref{sec:Lyp} we derive the basic equations for obtaining the quantum Lyapunov exponent, and solve explicitly the semi-classical case described by the Bohr-Sommerfeld quantization (Sec.~\ref{sec:LypBZ}). Numerical simulations demonstrating the analytical predictions are presented in Sec.~\ref{sec:numerical}.  Finally, we discuss the results in Secs.~\ref{sec:bound} and~\ref{sec:dis}, where the former focuses on the implications for the quantum bound on chaos. Technical details are given in four Appendices.

\section{Notations}
\label{sec:notation}
In the next Sections we discuss classical along quantum models. The mathematical language of the latter consists of different  objects and operations, such as matrices and tensors. To facilitate the reading we now specify the different notations that we  use throughout the paper:
\begin{enumerate}
\item Superoperators-- which operates on matrices and return matrices---  are denoted by calligraphic letters, e.g., $\mathcal{J}$.
\item The superoperators (tensors of rank 4) act on operators (matrices) according to the following  definition and notation  
$$
(\mathcal{F}\odot O)_{nn'}\equiv\sum_{n_2,n_1}\mathcal{F}_{nn'n_1n_2}O_{n_2n_1}.
$$
\item
Matrix multiplication operates as usual
$$
\begin{pmatrix}
\mathcal{A}\odot & \mathcal{B}\odot\\
\mathcal{C}\odot & \mathcal{D}\odot
\end{pmatrix}
\begin{pmatrix}
X \\
Y
\end{pmatrix}=
\begin{pmatrix}
\mathcal{A}\odot X +\mathcal{B}\odot Y \\
\mathcal{C}\odot X +\mathcal{D}\odot Y
\end{pmatrix}
$$
\item We work in the Heisenberg picture where operators are time-dependent. The subscript $0$ refers to the initial value.
\end{enumerate}

\section{Quantum Lyapunov exponent and the quantum tangent space}
\label{sec:tangent}

In Classical Mechanics, the basic measure of chaos is  the divergence  of two initially close trajectories in the phase space, $ \vec{x} \equiv (\vec{q},\vec{p})$. An exponential separation--- defining the Lyapunov exponent--- signifies chaos. 
The standard procedure to calculate Lyapunov exponents is by considering the tangent space, which describes the evolution of the distance between a pair of infinitesimally separated trajectories $\vec{u}\equiv \vec{x}(t)-\vec{x}'(t)$. It is dictated by the linear relation $\dot{\vec{u}}=M(\vec{x}(t))\vec{u}$, where the matrix $M$ contains  the second derivatives of the Hamiltonian evaluated along a reference trajectory $\vec{x}(t)$ in  phase-space (see e.g., Appendix~\ref{app:classical}). The largest Lyapunov exponent is then defined by 
\begin{multline}
\lambda_{\rm cl}\equiv \lim_{t\rightarrow \infty}\lim_{||u(0)||\rightarrow 0} \frac{1}{2t}\ln \frac{||u||^2}{||u(0)||^2}=\\
\lim_{t\rightarrow \infty} \ln \left (\frac{\lambda_{\max} \left[(\mathcal{T} e^{\int^t M(t')dt'})^T  \mathcal{T} e^{\int^t M(t')dt'} \right]}{2t}\right),
\label{eq:Lypdef}
\end{multline}
where $\lambda_{\max}[A]$ is the maximal eigenvalue of $A$, and $\mathcal{T}$ denotes time ordering.

Because we will be interested in the relation with quantum mechanics, it is natural to define the tangent space dynamics  with Poisson brackets. They  satisfy a chain rule: for any pair of conjugate variables, $(p,q)$ and some function $F(q,p)$ we have
\begin{equation}
\{F,p_0\}=-\{p,p_0\}\{F,q\}+\{q,p_0\}\{F,p\},
\label{eq:chainpq}
\end{equation}
where $(q_0,p_0)$ corresponds to the value at time zero. In Eq.~\eqref{eq:chainpq} we also make use of the fact that the the Poisson brackets are canonical invariants; see explicit derivation in Appendix~\ref{app:classical}. Based on this chain rule, from the Hamilton equations one finds 
\begin{equation}
\frac{d}{dt}
\begin{pmatrix}
\{p,q_0\}\\
\{q,q_0\}
\end{pmatrix}=
\begin{pmatrix}
\{\{H,p\},q\}&-\{\{H,p\},p\}\\
\{\{H,q\},q\}&-\{\{H,q\},p\}	
\end{pmatrix}\begin{pmatrix}
\{p,q_0\}\\
\{q,q_0\}	
\end{pmatrix}.
\label{eq:ODEPos}
\end{equation}
The matrix that appears in Eq.~\eqref{eq:ODEPos} is exactly $M(t)$ that governs the dynamics of the displacement vector $\vec{u}$ in the tangent space. The initial condition for Eq.~\eqref{eq:ODEPos} is the vector $(0,-1)$. One should also consider the other set of Poisson brackets $\{\cdot,p_0\} $. However, these are decoupled from the set $\{\cdot,q_0\}$ and satisfy the same linear relation with an initial condition $(1,0)$. Therefore it is sufficient to study the dynamics dictated by the matrix $M(t)$ for any initial condition.

Let us now turn to quantum mechanics. It is natural to implement quantization  by replacing Poisson brackets by commutators $\{\cdot,\cdot\}\rightarrow i\hbar[\cdot,\cdot]$, {\em but now we have to take care of factor orderings}. One thus finds that the dynamics of an OTOC is dictated by a linear relation 
\begin{equation}
i\hbar\frac{d}{dt}
\begin{pmatrix}
[A,B_0]\\
[B,B_0]\\
\end{pmatrix}
=\begin{pmatrix}
\mathcal{K}_{1}\odot&\mathcal{K}_{2}\odot\\
\mathcal{K}_{3}\odot&\mathcal{K}_{4}\odot\\
\end{pmatrix}
\begin{pmatrix}
[A,B_0]\\
[B,B_0]\\	
\end{pmatrix},
\label{eq:ODEC}
\end{equation}
where, for example, $\mathcal{K}_1$ and $\mathcal{K}_2$ come from $[[A,H],B_0]$.  Equation (\ref{eq:ODEC}) may be seen as a Lindbladian
expression for the tangent-space evolution.  
In order to prove the  relations playing the role of the chain-rule for the commutators, we use the fact that for any analytic function 
$ g(A) = \sum_n d_n A^r$
  we have  $[g(A),B]=\mathcal{S}_g\odot [A,B],$ with the  superoperator $\mathcal{S}_g$ acting as
\begin{equation}
\mathcal{S}_g\odot [X] = \sum_{n=0}^{\infty} \sum_{r=1}^{n}  d_n  A^{r-1}  X A^{n-r}.
\end{equation}
The superoperation is just a combination of left and right matrix multiplications, see Appendix~\ref{app:chain} for more details. As an example, if we take $A=p$, $B=q$ and $H=p^2+q^4$, then we get
\begin{multline}
[[A,H],B_0]=[[p,p^2+q^4],q_0]=-4i\hbar [q^3,q_0]=\\
-4i\hbar\left([q,q_0]q^2+q[q,q_0]q+q^2[q,q_0] \right),  
\end{multline}

Note that the initial conditions for the ODE in Eq.~(\ref{eq:ODEC}) is $([A_0,B_0],0)$, however as in the classical case we can also consider commutators of the form $[\cdot,A_0]$, for which we have the initial condition $(0,[B_0,A_0])$. Therefore, in principle, it is sufficient to consider the above matrix of superoperators for general initial conditions, although one should bare in mind that the magnitude of these must be bounded, as, e.g.,  $|[p_0,q_0]|=\hbar$.

As discussed in the Introduction, a quantum Lyapunov exponent $\lambda_{Q}$ can be defined via $\langle[A(t),B_0]^2\rangle\sim e^{2\lambda_Q t}$. In the current paper we mostly focus on a microcanonical version, taking $\bra{\psi_0}[A(t),B_0]^2\ket{\psi_0}$ with an initial eigenfunction $\ket{\psi_0}$. The canonical version is discussed in Sec.~\ref{sec:bound}.

Equation~\eqref{eq:ODEC} gives us a convenient framework to explore the growth of the OTOC, and its analogy to the classical Lyapunov separation. 
Ideally, we should compute the average of the {\em logarithm} of the squared commutator. For simplicity, more often the average of the squared commutator itself
is computed, which thus constitutes an {\em annealed} average: this is what we shall do in this paper.
We focus on a general class of models: Integrable systems which are weakly perturbed by an external noise.

\section{One dimensional model}
\label{sec:model}
Our goal is to understand the Lyapunov exponent of a quantum integrable Hamiltonian, $H_{\ing}$, which is weakly perturbed by additive noise. 
It turns out that the mechanism whereby chaos is induced by noise is already well represented by a system of one degree of freedom.
Transforming
to action-angle variables it reads:
\begin{equation}
H(N,e^{i\Theta})=H_{\ing}(N)+\epsilon^{1/2}\eta(t)G(N,e^{i\Theta}), 
\label{eq:1D}
\end{equation}
where the action-like operator $N$ counts the energy level number of $H_{\ing}$, the operator $e^{i\Theta}$ satisfies the commutation relation $[N,e^{i\Theta}]=e^{i\Theta}$, and $\eta$ is a Gaussian white noise. Working with the operator $e^{i\Theta}=\cos\Theta+i\sin\Theta$ allows us to easily relate the quantum problem to the classical action-angle variables, while avoiding an explicit use of the problematic phase-opertor $\Theta$. The operator $e^{i\Theta}$ itself suffers from some perplexing properties~\cite{carruthers1968phase,Nieto1993} such as $e^{i\Theta}e^{-i\Theta}\neq e^{-i\Theta}e^{i\Theta}$,  but this will pose no problem. Apart from the commutation relation stated above, we have $e^{im\Theta}\ket{n}=\ket{n+m}$. Thus, alternatively one can work with the more familiar ladder operator $a^{\dagger}$, where the number operator reads $N=a^{\dagger}a$.

The model in Eq.~\eqref{eq:1D} may involve any functional form for $H_{\ing}$ and $G$. Nevertheless, in this paper we start with a concrete class of classical models--- a particle in a power potential weakly perturbed by a random field--- and its quantum counterpart. The classical Hamiltonian reads: 
\begin{equation}
H_{\rm cl}=H_{{\rm cl} ,\ing}+\epsilon^{1/2} q \eta(t)   \qquad , \qquad H_{{\rm cl} ,\ing}=\frac{p^2}{2m}+\alpha q^\nu,
\label{eq:Hclass}
\end{equation}
where $m$ is the mass of the particle and $0<\nu<\infty$. Being one-dimensional, $H_{{\rm cl}, \ing}$ is integrable, and the action coordinate can be calculated explicitly~\cite{Carinena_etal1993}, giving 
\begin{equation}
{\rm classical:}\,\,\, H_{\rm cl}=KI^{\gamma}+\epsilon^{1/2} q(I,e^{i\Theta})\eta(t),
\label{eq:Hcl}
\end{equation} 
where $\gamma=2\nu/(2+\nu)$, $q(I,e^{i\Theta})\propto I^{2/(2+\nu)}$, and $K$ is a function of $m$, $\alpha$ and $\nu$ given in Eq.~\eqref{eq:HclI}. The form of the quantum version can be deduced directly from of Eq.~\eqref{eq:Hclass} by using scaling arguments: we look for a rescaling of momentum $p\rightarrow b p$,  such that coordinate rescales as  $q\rightarrow (\hbar/b) q$. We find (see Appendix~\ref{app:quantization})
\begin{multline}
{\rm quantum:}\,\,\,  H=\omega_0\hbar \tilde{H}(N,e^{i\Theta})= \\
\omega_0\hbar\left[\tilde{H}_{\ing}(N)+\tilde{\epsilon}^{1/2} \tilde{q}(N,e^{i\Theta}) \eta(t)\right],
\label{eq:HQ}
\end{multline}
where $\omega_0=\alpha^{1-\frac{\gamma}{2}}m^{-\frac{\gamma}{2}}\hbar^{\gamma-1}$ has units of frequency, and $\tilde{\epsilon}=\epsilon/(m^{1-\gamma}\alpha^{2-\gamma}\hbar^{2\gamma-1})$ is adimensional. In Eq.~\eqref{eq:HQ} we have also rescaled time by $\omega_0$, $\eta(t)\rightarrow \omega_0^{1/2}\eta(t)$,  such that {\em all the quantities in the square brackets are separately a-dimensional}. 

The Bohr-Sommerfeld approximation   of the time-independent part can be inferred by inserting the quantization relation $I=\hbar N$ in Eq.~\eqref{eq:Hcl}, which gives $\tilde{H}_{\ing}(N)=N^{\gamma}$ and $\tilde{q}\propto N^{\mu}$ with $\mu=(2-\gamma)/2$~\cite{Carinena_etal1993}\footnote{A proof that the Bohr-Sommerfeld approximation is valid for the case of one-dimensional power potential can be found in Ref.~\cite{Voros1994}.
}.  This approximation  is used to derive the  semi-classical Lyapunov in Sec.~\ref{sec:LypBZ}.

The quantum Lyapunov exponent should be understood for two copies that evolve and `feel' the same noise realization. In other words, if alternatively we considered a Loschmidt echo, then we would do the following~\cite{Kurchan2018}: evolve the system forward up to time $T$ with $H(N,e^{i\Theta},\eta(t))$, and then backward from $T$ to $2T$ with $-H(N,e^{i\Theta},\eta(2T-t))+\delta H$, where $\delta H$ is some small perturbation, and the noise  for the backward evolution  is the {\it time-reversed} of the one for the forward one. Next we summarize the chaotic behavior of the classical model.

\section{Summary of the classical case}
\label{sec:classical}

The Lyapunov exponent of a general classical integrable model perturbed by noise was derived in Ref.~\cite{Lam&Kurchan2014}.  We now briefly present the analysis  of this derivation as applied to the Hamiltonian in Eq.~\eqref{eq:Hcl}.
 {\em The different steps of the derivation shall be followed as closely as possible when we treat the quantum case in Sec.~\ref{sec:Lyp}.}

\subsection{Classical tangent space dynamics}

The motion in the tangent space is dictated by the Langevin dynamics
\begin{multline}
\frac{d}{dt}\begin{pmatrix}
u_I\\
u_{\Theta}
\end{pmatrix}=\left[\begin{pmatrix}
0 & 0\\
H''_{\rm cl,\ing} & 0
\end{pmatrix}
+\epsilon^{1/2}\eta(t)
K(t)\right]
\begin{pmatrix}
u_I\\
u_{\Theta}
\end{pmatrix}\approx\\
\approx \begin{pmatrix}
0 & \epsilon^{1/2}\left(\partial^2_{\Theta}q(I_0,e^{i\Theta})\right)\eta(t)\\
H''_{\rm cl,\ing}(I_0) & 0
\end{pmatrix}
\begin{pmatrix}
u_I\\
u_{\Theta}
\end{pmatrix},
\label{eq:varCl}
\end{multline}
where $'$ refers to derivative with respect to the action variable $I$, and $K(t)$ is a matrix which depends on the second derivatives of the perturbation. The structure of the final matrix results from power-counting in $\epsilon$, after assuming that the matrix is evaluated along unperturbed reference trajectory (along which the tangent space is measured) $I(t)=I_0$, $\Theta(t)=\Theta_0+H'_{\rm cl,\ing}(I_0)t$. 
{\em An important remark: at this point one can scale out $\epsilon$ by rescaling $u_{I} \rightarrow u_I$, $u_{\Theta} \rightarrow \epsilon^{-1/3}u_{\Theta}$ and $t \rightarrow \epsilon^{-1/3} t$, and thus immediately reach  the conclusion that $\lambda_{\rm cl}\propto \epsilon^{1/3}$}. However, we shall continue without this rescaling since it is not crucial for the purpose of this section.

The  Fokker-Planck equation, describing the evolution of the probability distribution of $(u_I,u_{\Theta})$, reads
\begin{widetext}
\begin{equation}
\frac{\partial P(u_I,u_{\Theta})}{\partial t}=\left\{-H''_{\rm cl,\ing}u_I \frac{\partial}{\partial u_{\Theta}}+\right.
\left.  \epsilon \left(\partial^2_{\Theta}q(I_0,e^{i\Theta})\right)u^2_{\Theta} \frac{\partial^2}{\partial u^2_{I}} \right\} P(u_I,u_{\Theta}).
\end{equation}
{\em This equation is homogeneous }, thanks to this
 we can derive a close set of ODEs describing averages over the noise of quadratic quantities:
\begin{equation}
\frac{d}{dt}
\begin{pmatrix}
\langle u^2_I \rangle\\
\langle u^2_{\Theta} \rangle \\
\langle u_I u_{\Theta} \rangle
\end{pmatrix}
=\begin{pmatrix}
0 & \epsilon \left(\partial^2_{\Theta}q(I_0,e^{i\Theta})\right)^2 & 0\\
0 & 0 & 2 H''_{\rm cl,0} \\
H''_{\rm cl,0} & 0 & 0
\end{pmatrix}
\begin{pmatrix}
\langle u^2_I \rangle\\
\langle u^2_{\Theta} \rangle \\
\langle u_I u_{\Theta} \rangle
\end{pmatrix}.
\label{eq:setC}
\end{equation}
\end{widetext}
This set of equations describes the evolution of  the annealed Lyapunov exponent. Under the assumption that rotation around the torus is faster than the Lyapunov time,  $\lambda_{\rm cl}\ll H'_{\rm cl,\ing}(I_0)$, one can  average over the angle $\Theta$, 
defining the important parameter for what follows
\begin{equation}
 \bar{q} \equiv \sqrt{ \langle \left(\partial^2_{\Theta}q(I,e^{i\Theta})\right)^2 \rangle_{\Theta}}.
\label{qbar}
\end{equation}
Replacing the term by its average may be understood as a first term in a Magnus expansion (see Appendix~\ref{app:Magnus}). The resulting $3\times3$ eigenvalue problem gives the annealed Lyapunov
\begin{equation}
2\lambda_{\rm cl}=2^{2/3}\epsilon^{1/3} \left(H''_{\rm cl,\ing}\right)^{2/3} \bar{q}^{2/3}(I_0).
\label{eq:Lypcl}
\end{equation}

The above derivation relies on the assumption of weak perturbation--- the diffusion time for action variables  is shorter than the Lyapunov time
\begin{equation}
\lambda^{-1}_{\rm cl} \ll \frac{I^2_0}{ \epsilon \bar{q}^2(I_0)}.
\label{eq:weakC}
\end{equation}
Since $\lambda_{\rm cl}\sim \epsilon^{1/3}$, this inequality holds for small enough $\epsilon$. 
The  $\lambda_{\rm cl}\propto\epsilon^{1/3}$ scaling was already found in the context of motion along a stochastic magnetic field~\cite{Rechester_etal1979}, and in the theory of products of random matrices~\cite{Anteneodo2001,Vallejos2012}.

\subsection{ Classical prescrambling time}
\label{sec:ClPreSc}

In order to better understand the influence of the quantum dispersion of the initial condition, we need to see first what happens
classically when the initial separation of trajectories is finite.
 Let us thus consider two initial nearby  trajectories at a non-infinitesimal initial separation $(u_{I,0},u_{\Theta,0})$. There is an initial time window, $[0,t_b]$, within which the $u_I$ stays small, while their angular separation grows {\em ballistically} 
 $u_{\Theta}(t_b)\approx u_{\Theta,0}+H''_{\rm cl,\ing}(I_0)u_{I,0}t_b$. This is followed by the Lyapunov regime, where the two trajectories separate exponentially at a rate $\lambda_{\rm cl}$. In general, it is expected that the exponential growth starts after one Lyapunov time $t_b=\lambda^{-1}_{\rm cl}$ (we confirm this numerically  in Sec.~\ref{sec:numerical}), and
  is expected to saturate after some finite time, when
 the separation has grown to the size of the torus.  Na\"ively, one would think 
 that the saturation time $t_s$  may be estimated in the usual way, as the time $u_{\Theta,0} e^{\lambda_{\cl} t_s} \sim 2\pi$. However, this is not quite true. What happens
 is that it is not the initial time separation that is amplified by the exponential separation, but rather the separation {\em after the ballistic regime}, i.e.
  $u_{\Theta}(t_b) \sim u_{\Theta}(\lambda_{\cl}^{-1})$ we obtained above.
 We hence have:
\begin{equation}
t_s\equiv \lambda^{-1}_{\rm cl} \log \left(\frac{2\pi}{u_\Theta(t_b)}\right) \sim \lambda^{-1}_{\rm cl} \log \left(\frac{2\pi}{u_\Theta(\lambda^{-1}_{\rm cl})}\right)
\label{eq:ts}
\end{equation}

Finally, let us check that during these times the diffusion of the action is small. At the saturation time
\begin{equation}
u_I(t_s)=u_I(\lambda^{-1}_{\rm cl})e^{\lambda_{\rm cl}t_s}=
\frac{2\pi u_I(\lambda^{-1}_{\rm cl})}{u_{\Theta}(\lambda^{-1}_{\rm cl})}.
\end{equation}
Plugging in the expression for $u_{\Theta}(\lambda^{-1}_{\rm cl})$, and the solution for $\lambda_{\rm cl}$ from Eq.~\eqref{eq:Lypcl} we find
\begin{equation}
u_I(t_s)=\frac{2\pi}{H''_{\rm cl,\ing}(I_0)\lambda^{-1}_{\rm cl}}=\pi\left(\frac{\epsilon \bar{q}^2}{H''_{\rm cl,\ing}}\right)^{1/3} \ll I_0,
\end{equation}
where the inequality comes from the assumption of weak perturbation in Eq.~\eqref{eq:weakC}. Thus, we confirm that small perturbation that corresponds to very slow diffusion  results in a  Lyapunov separation with small projection along the action coordinates. All of these results are verified numerically in Sec.~\ref{sec:numerical}, see Fig.~\ref{fig:ClLyps}.

\section{Outline of main results for the quantum case}
\label{sec:summ}

Let us now consider the quantum model in Eq.~\eqref{eq:HQ}. The model depends on  {\it two a-dimensional numbers}: the typical energy level $n_0$ and the adimensional perturbation strength $\tilde{\epsilon}$. 
In the current section we provide general arguments for the behavior of the quantum model, and outline our main results with a scheme in Fig.~\ref{fig:scheme}. 

\subsection{Quantum tangent space dynamics}

As we show in Sec.~\ref{sec:Lyp}, it is sufficient to focus on the dynamics of two operators, $C^{\Theta}=[e^{i\Theta},A_0]e^{-i\Theta}$ and $C^{N}=i[N,A_0]$, with $A_0$ being some initial Hermitian operator. The time derivative of these operators follows a linear  super-operator  equations, which are the analogue of those of Eq.~\eqref{eq:varCl}, of the form
\begin{equation}
\begin{pmatrix}
\dot{C}^N \\
\dot{C}^{\Theta}	
\end{pmatrix}
=\left[
\begin{pmatrix}
0 & 0\\
\mathcal{L}\odot & i\mathcal{J}\odot
\end{pmatrix}+ \tilde \epsilon^{1/2}\eta(t)
\begin{pmatrix} 
\mathcal{M}\odot & \mathcal{F}\odot\\
\mathcal{N}\odot &\mathcal{K}\odot
\end{pmatrix}\right]
\begin{pmatrix}
C^{N}\\
C^{\Theta}	
\end{pmatrix},
\label{eq:varQ0}
\end{equation}
where $\mathcal{L},\mathcal{J},
\mathcal{M}, \mathcal{F},
\mathcal{N},\mathcal{K}$ are super-operators  easily obtained by using the `chain-rule' for commutators.

The quantum Lyapunov exponent is adimensionalized as: $\tilde {\lambda}_Q = \frac{\lambda_Q}{\omega_0}$, with $\tilde{\lambda}_Q$ defined by the (adimensional) time dependence of the OTOC  associated with the evolution generated by the adimensional Hamiltonian $H/(\omega_0\hbar)$:
\begin{equation}
C^2(t)\equiv \bra{\psi_0}C_{\Theta}^2\ket{\psi_0} \sim e^{2\tilde{\lambda}_Q t}.
\label{eq:OTOC}
\end{equation}
The initial state, $\ket{\psi_0}=\sum_n c_n \ket{n}$, is assumed to be concentrated around some $n_0$, here we take directly $\ket{\psi_0}= \ket{n_0}$.

The super-operator $\mathcal{J}$ comes from factor reorderings, and thus vanishes in the classical case. When this is so, we may rescale $C^N \rightarrow C^N$, $C^\Theta \rightarrow \tilde \epsilon^{-1/3}C^\Theta$ and $t \rightarrow \tilde \epsilon^{-1/3} t$, and conclude that $\mathcal{M}, \mathcal{N},\mathcal{K}$ may be neglected for small $\tilde \epsilon$, and that the Lyapunov exponent $\tilde \lambda_Q$ scales like $\tilde{\epsilon}^{1/3}$. In the quantum case $\mathcal{J}\neq 0$, and time cannot be rescaled with $\tilde{\epsilon}$.

As in the classical case, we assume weak perturbation,  such that there is a negligible diffusion of energy levels during one Lyapunov time, $t_d\gg t_{\rm Lyp}$. In adimensional units this reads:
\begin{equation}
\tilde{\epsilon} \overline{q}^{2}(n_0)\ll  n_0^{3/2} H''(n_0),
\label{eq:weak}
\end{equation}
where the energy  diffusion rate is approximated hereafter as $n_0 \sim\tilde{\epsilon}\overline{q}^{2}(n_0)   t_d$, and $\overline{q}^{2}(n_0)$ is defined in Eq.~(\ref{qbar}). For large values of $\epsilon$, the $\epsilon^{1/3}$ scaling breaks down, e.g.,~\cite{Chirikov1979,Rozenbaum_etal2017}.


\subsection{Quantum prescrambling time}

Let us now see how what we have learned about saturation times in classical case (Sec.~\ref{sec:ClPreSc}) affects the quantum picture. Following Refs.~\cite{Chirikov_etal1988,Berman&Zaslavsky1978}, one shall imagine the semiclassical spreading of an initial wavepacket. The size of the packet prior to the exponential growth includes an 
initial  ballistic regime, $\ell(t_b)=\ell_0+v_0 t_b \sim \ell_0+v_0/\lambda_{cl}$. Now, the uncertainty principle implies that both $\ell_0$ and $v_0$ are finite, for example considering a coherent state
as an initial packet. Quantum effects thus saturate the exponential growth once the wavepacket spreads throughout the torus, $\lambda_{\cl} t \sim -\log(\ell_0+v_0 /\lambda_{\cl})$.

Quantum mechanics thus acts in two forms: for large Lyapunov exponents it is the initial wavepacket that spreads, just as in the usual Ehrenfest time estimate -- {\em only
that here it concerns spreading over the prethermalization space (the quantum counterpart of the torus)  rather than over the entire phase-space}.  For small Lyapunov exponents, as we shall have when the perturbation is weak, the ballistic time is long, and hence the quantum spread of initial velocities  could even reach 
prescrambling. These two effects thus limit separately the conditions under which there is a Lyapunov time at all: if the prescrambling time is of the order of the (classical) Lyapunov time, then the Lyapunov regime is finished before it starts.

\subsection{Overview of the results}

\begin{figure}
\centerline{\resizebox{0.45\textwidth}{!}{\includegraphics{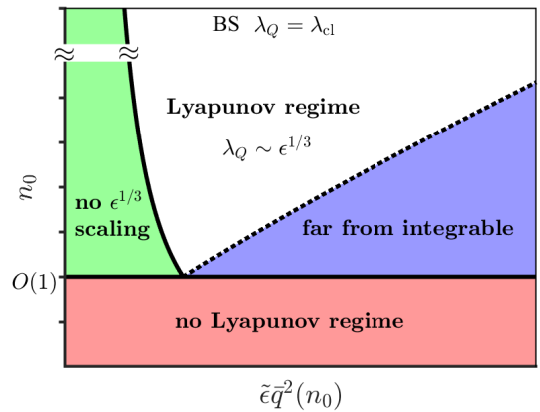}}}
\caption[]
{ A scheme summarizing our findings for the OTOC dynamics in quantum quasi-integrable systems described by Eq.~\eqref{eq:HQ}. The analysis assumes weak perturbation, Eq.~\eqref{eq:weak}, thus excluding the blue regime. When $\mathcal{J}$ in Eq.~\eqref{eq:varQ0} is negligible, a Lyapunov exponent which scales as $\tilde{\epsilon}^{1/3}$ is predicted based on scaling arguments (white area; this is verified in Figs.~\ref{fig:Lyps2ab} and~\ref{fig:Lyps}). A semiclassical derivation based on the BS for $n_0$ large enough is given in Sec.~\ref{sec:LypBZ}. From the discussion on the prescrambling times in Sec.~\ref{sec:summ}, we can predict that the Lyapunov regime vanishes at low quantum numbers (red area;  supported by Fig.~\ref{fig:Lyps2ab}a and~\ref{fig:Lyps2}), and at very low perturbation (green area; supported by Fig.~\ref{fig:Lyps2ab}b and~\ref{fig:Lyps}).
}
\label{fig:scheme}
\end{figure}

A summary of our main conclusions is given schematically in Fig.~\ref{fig:scheme}. The different regimes are based on the physical arguments presented in the current section, the complete analysis in Sec.~\ref{sec:Lyp}, and its verification with numerical simulations in Sec.~\ref{sec:numerical}. We indicate explicitly what is the numerical evidence/analytical derivation from which we conclude each part of the diagram.

We now move to the calculation of the quantum Lyapunov exponent.

\section{Analytical description for the OTOCs dynamics and the quantum Lyapunov exponent}
\label{sec:Lyp}

The goal of this Section is to calculate the growth rate of the square of a commutator, $\bra{\psi}C^2\ket{\psi}$.  The most natural thing is to consider a linear combination of states around some $|n_0\rangle$, corresponding to a wavepacket in the angular variables. In fact, we shall use a single state
 $\ket{\psi}=\ket{n_0}$, and check that the results correspond to those of a packet. 
The state  $|n_0\rangle$ is spread over {\em all} angles, which seems at odds with the interpretation of the Lyapunov regime as
the time during which a packet has not spread. It should be born in mind, however, that $|n_0\rangle$ has oscillations in $\Theta$ of length $1/n_0$, and these dephase completely in times similar to that of a wave packet (see inset of Fig.~\ref{fig:Lyps2}). Finally, the implications of our results for thermal averages  are discussed in the last section, Sec.~\ref{sec:bound}.

The current section is organized as follows: The equations for the tangent space dynamics and the annealed Lyapunov exponent are derived in Sec.~\ref{sec:OTOC} and Sec.~\ref{sec:LypAnn} respectively. Then, in Sec.~\ref{sec:simpl} we treat further a slightly simplified model and solve explicitly the semi-classical limit in Sec.~\ref{sec:LypBZ}.

\subsection{Evolution of the OTOCs}
\label{sec:OTOC}

We focus on the dynamics of two operators 
\begin{equation}
C^{\Theta}\equiv [e^{i\Theta},A_0]e^{-i\Theta},\,\,\,
C^{N}\equiv i[N,A_0],
\label{eq:OTOCS}
\end{equation}
with $A_0$ being some initial Hermitian operator. The choice of normalization of the first commutator with $e^{-i\Theta}$ is analogous to the classical counterpart, there it compensates for the fact that we are working with non-canonical variables $N$ and $e^{i\Theta}$ (see Appendix~\ref{app:classical}). The normalization also guarantees that $C^{\Theta}$ is Hermitian. We find that this choice of OTOCs facilitates the analytic derivation, however, the same dynamics is expected for other operators such as $[\cos\Theta,A_0]$.

Let us look at the time derivatives of the above two operators (recall that time is rescaled by $\omega_0$):
\begin{eqnarray} 
\dot{C}^{N}&=&[[N,\tilde{H}_{\ing}(N)+\tilde{\epsilon}^{1/2} \tilde{q}(N,e^{i\Theta}) \eta(t)],A_0]\sim O(\tilde{\epsilon}^{1/2}),\nonumber\\
\dot{C}^{\Theta}&=&-i[[e^{i\Theta},\tilde{H}],A_0]e^{-i\Theta}-i[e^{i\Theta},A_0][e^{-i\Theta},\tilde{H}].
\label{eq:Cdotraw}
\end{eqnarray}
The first term on the right hand side of Eq.~\eqref{eq:Cdotraw} can be written as
\begin{multline}
-i[[e^{i\Theta},\tilde{H}]e^{-i\Theta},A_0]+i[e^{i\Theta},\tilde{H}][e^{-i\Theta},A_0]\\
=-i[[e^{i\Theta},\tilde{H}]e^{-i\Theta},A_0]-i[e^{i\Theta},\tilde{H}]e^{-i\Theta}C^{\Theta},
\end{multline}
where we have applied the relations $[A,C]B=[AB,C]-A[B,C]$ and $
[e^{-i\Theta},\cdot]=-e^{-i\Theta}[e^{i\Theta},\cdot]e^{-i\Theta}$. 
 In addition, using this last relation, the second term in Eq.~\eqref{eq:Cdotraw} can be rewritten as
\begin{equation}
-i[e^{i\Theta},A_0][e^{-i\Theta},\tilde{H}]=
iC^{\Theta}[e^{i\Theta},\tilde{H}]e^{-i\Theta}.
\end{equation}
In summary we have
\begin{equation}
\dot{C}^{\Theta}=i[\Omega(N),A_0]+i[\Omega(N),C^{\Theta}]+O(\tilde{\epsilon}^{1/2}),
\label{eq:CTdot}
\end{equation}
where $\Omega(N)\equiv-[e^{i\Theta},\tilde{H}_{\ing}(N)]e^{-i\Theta}=\tilde{H}_{\ing}(N)-\tilde{H}_{\ing}(N-1)$.

The corresponding equations, which are equivalent to those of the tangent space in classical mechanics, are of the form
\begin{equation}
\begin{pmatrix}
\dot{C}^N \\
\dot{C}^{\Theta}	
\end{pmatrix}
=\left[
\begin{pmatrix}
0 & 0\\
\mathcal{L}\odot & i\mathcal{J}\odot
\end{pmatrix}
+
\tilde{\epsilon}^{1/2}\eta(t) \begin{pmatrix}
\mathcal{M}\odot & \mathcal{F}\odot\\
\mathcal{N}\odot & \mathcal{K}\odot
\end{pmatrix}
\right]
\begin{pmatrix}
C^{N}\\
C^{\Theta}	
\end{pmatrix}.
\label{eq:varQ}
\end{equation}

Let us now write explicitly the operators $\mathcal{J}$ and $\mathcal{L}$, which depend only on the integrable part $\tilde{H}_{\ing}$. We work in the eigenbasis $\ket{n}$ and denote the period of the torus 
\begin{equation}
\omega_n\equiv E_n-E_{n-1},
\end{equation}
with $E_n$  the energy levels of the integrable model.  

$\bullet$ From Eq.~\eqref{eq:CTdot} we have for $\mathcal{J}$:
\begin{equation}
(\mathcal{J}\odot C)_{nn'}=i(\omega_{n}-\omega_{n'})C_{nn'}\equiv i \; j(n,n')C_{nn'},
\label{eq:supJ}
\end{equation}

$\bullet$ The superoperator $\mathcal{L}$ can be represented  as a sum of left and right matrix multiplication, using the relation $[N^s,A_0]=\sum^s_{r=1} N^{s-1} [N,A_0]N^{s-r}$, alternatively, we can use the relation in Eq.~\eqref{eq:relation1} which leads to:
\begin{equation}
(\mathcal{L}\odot C)_{nn'}=\frac{\omega_{n}-\omega_{n'}}{n-n'}C_{nn'}\equiv l(n,n')C_{nn'}.
\label{eq:supL}
\end{equation}
 In the semiclassical limit, when the density of levels is high,  $n\sim n' \gg 1$, one has that $l(n,n') \rightarrow \partial^2 H/\partial n^2$ (see Sec.~\ref{sec:LypBZ}).

Before we proceed with the equations for the annealed Lyapunov exponent, we briefly discuss the integrable case.

\subsubsection{Integrable case}
\label{sec:integrable}

When there is no external noise, $\tilde{\epsilon}=0$, the dynamics of the OTOC follows
\begin{equation}
\begin{pmatrix}
\dot{C}^N \\
\dot{C}^{\Theta}	
\end{pmatrix}
=
\begin{pmatrix}
0 & 0 \\
\mathcal{L}\odot & i\mathcal{J}\odot
\end{pmatrix}
\begin{pmatrix}
C^{N}\\
C^{\Theta}	
\end{pmatrix}.
\label{eq:varQint}
\end{equation}
The above dynamics cannot yield an  exponential growth for the OTOC. Only $C^{\Theta}$ may grow exponentially, but since $i\mathcal{J}=i[\Omega(N),\cdot]$ we get
$C^{\Theta}(t) = e^{-i\Omega t} C^{\Theta}_0 e^{i\Omega t} $, an oscillatory term.

\subsection{Annealed Lyapunov exponent}
\label{sec:LypAnn}

The derivation which follows is done along the lines of the classical problem that was addressed in Ref.~\cite{Lam&Kurchan2014} and briefly discussed in Sec.~\ref{sec:classical}. We can do this since the formulation of the two problems is the same, the quantum case is just in higher number (infinite) of degrees of freedom: the variables $C_{nn'}$ can be thought as vectors and accordingly the superoperators can be thought as matrices. Nevertheless, there should be differences which come from quantum mechanics. 

The matrix of superoperators in Eq.~\eqref{eq:varQ} has elements that depend on time through the operators $N$ and $e^{i\Theta}$, which evolve according to the full perturbed Hamiltonian  in the Heisenberg picture. However, we can assume that the evolution is well approximated by the evolution unperturbed by noise, and the effect of noise is only important at the level of the tangent space. Note that the
same approximation is assumed in the classical case. For small $\tilde{\epsilon}$ (Eq.~\eqref{eq:weak}) the perturbation then gives only small corrections and we have:
\begin{equation}
\mathcal{F}(N,e^{i\Theta})=\mathcal{F}(N_0,e^{i\tilde{H}_{\ing}(N_0)t}e^{i\Theta_0}e^{-i\tilde{H}_{\ing}(N_0)t})+O(\tilde{\epsilon}).
\end{equation}
  
Now, we can employ power-counting in $\tilde{\epsilon}$ to eliminate several components in the matrix of superoperators. If we assume the scaling  $t\rightarrow \tilde{\epsilon}^{-\alpha}t$, $C^N\rightarrow C^N$ and $C^{\Theta}\rightarrow \tilde{\epsilon}^{-\beta}C^{\Theta}$, together with the fact that for white noise $\eta(at)=a^{-1/2}\eta(t)$, we find
\begin{equation}
\dot{C}^N=\tilde{\epsilon}^{1/2-\alpha}\eta(t)\mathcal{M}\odot C^{N}+
\tilde{\epsilon}^{1/2-\alpha-\beta}\eta(t)\mathcal{F}\odot C^{\Theta} , 
\end{equation}
\begin{multline}
\dot{C}^{\Theta}=\left[\tilde{\epsilon}^{-\alpha+\beta}\mathcal{L}+\tilde{\epsilon}^{1/2-\alpha+\beta}\eta(t)\mathcal{N}\right]\odot C^{N} + \\
\left[ i\tilde{\epsilon}^{-\alpha}\mathcal{J}+\tilde{\epsilon}^{1/2-\alpha}\eta(t)\mathcal{K}\right]\odot C^{\Theta}.
\end{multline}
Then, $\tilde{\epsilon}\ll 1$ implies that $\mathcal{M}$ can be neglected with respect to $\mathcal{F}$, as well as $\mathcal{N}$ compared to $\mathcal{L}$, and $\mathcal{K}$ compared to $\mathcal{J}$. We have then:  
\begin{equation}
\begin{pmatrix}
\dot{C}^N \\
\dot{C}^{\Theta}	
\end{pmatrix}
=
\begin{pmatrix}
0& \tilde{\epsilon}^{1/2}\eta(t)\mathcal{F}(t)\odot\\
\mathcal{L}\odot & i\mathcal{J}\odot
\end{pmatrix}
\begin{pmatrix}
C^{N}\\
C^{\Theta}	
\end{pmatrix}.
\label{eq:SimpvarQ}
\end{equation}
We keep both $\mathcal{L}$ and $\mathcal{J}$, the latter is a factor-ordering term that disappears in the classical case. 

This is a Langevin equation satisfied by the components of the commutator.  Next, we may repeat the steps we followed in the classical case, deducing from the Langevin equation  \eqref{eq:SimpvarQ} a  Fokker-Planck equation satisfied by the components of the commutator (whose complete form is given in Eq.~\eqref{eq:FP0}). Using the homogeneity in the same way, we obtain a closed equation for the quadratic averages of components:
\begin{widetext}
\begin{eqnarray}
\frac{d}{dt}\langle C_{n n'}^{N} C_{m m'}^{N}\rangle &=&  \tilde{\epsilon}\sum_{\substack{n_3,n_4 \\ m_3,m_4}}\mathcal{F}_{n n' n_3 n_4}(t) \mathcal{F}_{m m' m_3 m_4}(t) \langle   C_{n_4 n_3}^{\Theta} C_{m_4m_3}^{\Theta}\rangle,
\nonumber\\
\frac{d}{dt}\langle C_{nn'}^{\Theta} C_{mm'}^{\Theta} \rangle &=& l(n,n') \langle  C_{n n'}^{N}C_{mm'}^{\Theta}\rangle
+l(m,m')\langle C_{nn'}^{\Theta}  C^{N}_{m m'} \rangle 
+i\,\left(j(n,n') + j(m,m')\right) \langle  C_{n n'}^{\Theta}C_{mm'}^{\Theta}\rangle,
\nonumber\\
\frac{d}{dt} \langle  C_{n n'}^N C_{m m'}^{\Theta} \rangle  &=& l(m,m') \langle C_{n n'}^{N}  C^{N}_{m m'} \rangle
+i\,j(m,m') \langle C_{n n'}^{N}  C^{\Theta}_{m m'} \rangle,
\label{eq:ODEraw3}
\end{eqnarray} 
\end{widetext}
where the functions $j(n,n')$ and $l(n,n')=l(n',n)$ appear in Eqs.~\eqref{eq:supJ} and~\eqref{eq:supL}. This is a closed set of ODEs for the averaged products of matrix elements. One can take a subset of these equations and hope that they will form a closed set. Next we consider a simplified model for which this can be done. In particular, we are interested in quantities of the form $\langle \sum_{n_1}C_{n n_1} C_{n_1 n}\rangle$ which correspond to evaluating expectations of square-commutators at a given eigenstate $\ket{n}$.

\subsection{Simplified model}
\label{sec:simpl}

We now simplify the problem by employing  $\tilde{q}(N,e^{i\Theta})=V(N)\cos\Theta+\cos\Theta V(N)$, a particular case of  the general model in Eq.~\eqref{eq:HQ}. The main characteristics of the solution should hold for other functional forms, that is, taking higher harmonics $\cos(2\Theta)$, $\sin(\Theta)$, etc. We thus treat the a-dimensional Hamiltonian:
\begin{equation}
\tilde{H}=\tilde{H}_{\ing}(N)+\tilde{\epsilon}^{1/2}\eta(t)\left(V(N)\cos(\Theta)+\cos(\Theta)V(N)\right).
\end{equation}
The OTOCs dynamics is dictated by Eq.~\eqref{eq:SimpvarQ} that contains the superoperators $\mathcal{L}$, $\mathcal{J}$, and $\mathcal{F}$. The former two are related to the integrable part and already given in Sec.~\ref{sec:OTOC}.

Let us calculate the superoperator $\mathcal{F}$, which corresponds to the term  proportional to $C^{\Theta}$ in the operation  $[[N,V(N)\cos\Theta+\cos\Theta V(N)],A_0]$. We shall thus consider
$$
[[N,\cos\Theta],A_0]=\frac{1}{2}(C^{\Theta}e^{i\Theta}+e^{-i\Theta}C^{\Theta}).
$$
Working in the eigenbasis of $N$, where $\bra{m}e^{i\Theta}\ket{n}=e^{i\omega_m t}\delta_{m,n+1}$ and 
$\bra{m}e^{-i\Theta}\ket{n}=e^{-i\omega_{n} t}\delta_{m,n-1}$, we find that
\begin{multline}
\mathcal{F}_{nn'kk'} =\left(e^{i\omega_{n'}t}\delta_{n,k'}\delta_{n'+1,k}+e^{-i\omega_n t}\delta_{n+1,k'}\delta_{n',k}\right)
\\
\times \frac{1}{2}\left(V(n)+V(n')\right).
\end{multline}

Next, we write down a set of ODEs for the dynamics of averages, according to Eq.\eqref{eq:ODEraw3}. Defining 
$F_{nn'}^{XY}\equiv \langle C^{X}_{nn'}C^{Y}_{n'n} \rangle$, with $X$ and $Y$ being $N$ or $\Theta$, we find:
\begin{equation}
\label{eq:ODE1}
\frac{d}{dt} F_{n n'}^{NN} = \sum_{\substack{n_1,n_2 \\ m_1,m_2}}\tilde{\epsilon}\mathcal{R}_{nn'm_1 m_2n_1n_2}(t) \langle C^{\Theta}_{n_2 n_1}C^{\Theta}_{m_2 m_1} \rangle,
\end{equation}
\begin{equation}
\label{eq:ODE2}
\frac{d}{dt}F^{\Theta\Theta}_{n n'} = 2l(n,n') \frac{F_{n n'}^{N\Theta}+F_{n n'}^{\Theta N}}{2}, 
\end{equation}
\begin{multline}
\frac{d}{dt} \left(\frac{F_{n n'}^{N\Theta}+F_{n n'}^{\Theta N}}{2}\right) = l(n,n') F_{n n'}^{NN} \\
-ij(n,n')  \frac{F_{n n'}^{N\Theta}-F_{n n'}^{\Theta N}}{2},\\
\end{multline}
\begin{equation}
\label{eq:ODE4} 
\frac{d}{dt} \left(\frac{F_{n n'}^{N\Theta}-F_{n n'}^{\Theta N}}{2}\right) = -ij(n,n')  \frac{F_{n n'}^{N\Theta}+F_{n n'}^{N\Theta}}{2},
\end{equation}
where $\mathcal{R}_{nn'kk'll'}(t)\equiv \mathcal{F}_{n n' l l'}(t) \mathcal{F}_{n' n k k'}(t)$, and we use the fact that  $j(n,n')=-j(n',n)$ and $l(n,n')=l(n',n)$. Finally, we preform time-averaging for $\mathcal{R}$ (see Appendix~\ref{app:Magnus}), that is, we drop all of the oscillating components. As in the classical case, this procedure refers to the assumption that the Lyapunov time is much longer than the periods around the torus. We find $
\overline{\sum\mathcal{R}_{nn'm_1 m_2n_1n_2}(t)} \langle C^{\Theta}_{n_2 n_1}C^{\Theta}_{m_2 m_1}\rangle
=
\frac{1}{4}\left(V(n)+V(n')\right)^2\left(F^{\Theta\Theta}_{n,n'+1}+F^{\Theta\Theta}_{n+1,n'}\right)
$, and thus Eq.~\eqref{eq:ODE1} can be replaced by
\begin{equation}
\frac{d}{dt} F_{n n'}^{NN} =  \frac{\tilde{\epsilon}}{4}\left(V(n)+V(n')\right)^2\left(F^{\Theta\Theta}_{n,n'+1}+F^{\Theta\Theta}_{n+1,n'}\right).
\label{eq:ODE1new}
\end{equation}
This closes the equations for the $F^{XY}$ functions. 

Following the classical case, the closed set of equations may be transformed to an equation for a single commutator:
\begin{widetext}
\begin{equation}
\frac{d^3}{dt^3} F_{nn'}^{\Theta\Theta} + 2l^2(n,n') (n-n')^2 \frac{d}{dt} F _{nn'}^{\Theta\Theta}- \frac{\tilde{\epsilon}}{4} 2l^2(n,n')  (V(n)+V(n'))^2 \left( F^{\Theta\Theta}_{n,n'+1}+F^{\Theta\Theta}_{n+1,n'}\right) =0,
\label{eq:EigPro0}
\end{equation}
\end{widetext}
where we substitute $j^2(n,n')=(n-n')^2l^2(n,n')$. 

Equation (\ref{eq:EigPro0}) is the key outcome of the calculation--- it describes the growth of the norm of matrix elements $F^{\Theta\Theta}_{nn'}\equiv |C^{\Theta}_{nn'}|^2$, in a simple model which tries to capture generic properties of  quasi-integrable systems. This third order ODE should be accompanied with initial conditions that correspond to some energy shell, as we describe below. Eq.~(\ref{eq:EigPro0}) should contain (a) the exponential growth of expectation values within the prescrambling time, and (b) their spreading along energy levels. One might also conjecture that this equation shall give (c) the saturation, i.e., the prescrambling time. In the current paper we focus on the exponential growth, providing an explicit result for the semi-classical limit. The other two dynamical properties are left for future considerations. 

We already learn something from Eq.~(\ref{eq:EigPro0}): if the term with the first time-derivative were absent, we could absorb $\tilde \epsilon$ into time and conclude that we have an $\tilde{\epsilon}^{\frac13}$ scaling of the Lyapunov exponent, just as in the classical case (see remark in Sec.~\ref{sec:classical}). The term linear in $\frac{d}{dt}$, originates from the factor reordering induced by $\mathcal{J}\odot C^{\Theta}=[\Omega(N),C^{\Theta}]$, and is of purely quantum origin.

\subsubsection{Initial conditions}

The definition of $C^{\Theta}$ and $C^{N}$ in Eq.~\eqref{eq:OTOCS} concerns the commutation of a time-evolving operator with some initial Hermitian operator $A_0$. We may choose $A_0=\ket{\psi_0}\bra{\psi_0}$ as a projection on an initial wavepacket concentrated around an eigenstate $\ket{n_0}$, or in the extreme case, just as $\ket{n_0}\bra{n_0}$. The corresponding commutators at time $t=0$ then read
$$
C^{\Theta}_0=\ket{n_0+1}\bra{n_0+1}-\ket{n_0}\bra{n_0},\qquad C^{N}=0.
$$
Therefore, the solution to Eq.~(\ref{eq:EigPro0}) shall be obtained for a given initial condition $F^{\Theta\Theta}_{nn'}(t=0)$ which is zero almost everywhere (this is a third order ODE, it has three initial conditions which are related to $F^{NN}$ and $F^{\Theta N}$ through Eqs.~\eqref{eq:ODE2}-\eqref{eq:ODE1new}).

\subsection{Semi-classical Limit: Bohr-Sommerfeld approximation}
\label{sec:LypBZ}

The aim of the current Section is to investigate Eq.~\eqref{eq:EigPro0} in the semi-classical limit, and show that it yields the classical Lyapunov exponent in Eq.~\eqref{eq:Lypcl}. In this limit we have the Bohr-Sommerfeld approximation $\tilde{H}_{\ing}(N)=N^{\gamma}$ and $V(N)=N^{\mu}$, where $\gamma$ and $\mu$ are related to the power-law potential; see discussion after Eq.~\eqref{eq:HQ}. This should be taken together with the limit $n\rightarrow\infty$ (going back to the dimension-full variables, this is equivalent to taking the limit $\hbar\rightarrow 0$ while fixing the energy $\sim (\hbar N)^{\gamma}$). 

First, the function $l(n,n')$ for a given $n\gg 1$ and $n'=n+Z$ reads
$$
l(n,n+Z)=\frac{n^{\gamma}-(n-1)^{\gamma}-(n+Z)^{\gamma}+(n+Z-1)^{\gamma}}{Z}.
$$
Since we are interested in operators which are localized in energy space, we are focusing on the limit of $Z\ll n$. Then we find 
$l(n,n+Z)=\gamma(\gamma-1) n^{\gamma-2}$, which is simply $\partial^2 \tilde{H}_{\ing}(n)/\partial n^2$. In addition, the expression $n^{\mu}+(n+Z)^{\mu}$ that appears in Eq.~\eqref{eq:EigPro0} is approximated as $2n^{\mu}$.
Next, let us discuss the term which is proportional to $(n-n')^2\frac{dF}{dt}$. Whenever this term is negligible, we can rescale time to find $F\sim e^{\tilde{\lambda}_Q t}$ with $\tilde{\lambda}_Q^3 \sim l^2 \tilde{\epsilon} n^{2\mu}$. Hence, we can drop this term self-consistently if $\tilde{\lambda}_Q Z^2\ll \tilde{\epsilon}n^{2\mu}$. This criterion is equivalent to 
\begin{equation}
|l(n,n+Z) Z|=|\omega_n-\omega_{n+Z}|\ll \tilde{\lambda}_Q.
\label{eq:Zcrit}
\end{equation}
As we discuss for the classical problem in Sec.~\ref{sec:classical} and demonstrate explicitly in the next Section: when the inequality in Eq.~\eqref{eq:Zcrit}, which  depends on the initial condition, is not satisfied we do not expect to see a Lyapunov regime.

In summary, the semi-classical limit refers to large energy $n_0\rightarrow\infty$ and not too small perturbation $Z_0\ll \tilde{\epsilon}n_0^{2\mu+\gamma-2}$.  Inserting those limits into Eq.~\eqref{eq:EigPro0}, summing over $n'$, and recalling that $\sum_{n'} F_{nn'}^{\Theta\Theta}=\bra{n}(C^{\Theta})^2\ket{n}=(C^{\Theta})^2_{nn}$, we find
\begin{multline}
\frac{d^3}{dt^3}(C^{\Theta})^2_{nn} = 2\tilde{\epsilon}\left(\gamma(\gamma-1) n^{\gamma-2}\right)^2n^{2\mu}
\times \\
\left( (C^{\Theta})^2_{nn}+(C^{\Theta})^2_{n+1,n+1}\right).
\label{eq:EigProbSC}
\end{multline}

Finally, assuming that the initial condition is concentrated around $n_0$, such that $(C^{\Theta})^2_{n_0+1,n_0+1}\sim (C^{\Theta})^2_{n_0,n_0}$, we can find the exponential growth described by Eq.~\eqref{eq:EigProbSC} with the ansatz 
$(C^{\Theta})^2_{n_0,n_0}\sim e^{2\tilde{\lambda}_Q t}$ to find the resulting a-dimensional quantum annealed Lyapunov:
\begin{equation}
2\tilde{\lambda}_Q = 2^{2/3} \tilde{\epsilon}^{1/3}(\gamma(\gamma-1) n_0^{\gamma-2})^{2/3} n_0^{2\mu/3}.
\label{eq:LypSC}
\end{equation}
We verify the correspondence between the semiclassical Lyapunov and the classical one in Eq.~\eqref{eq:Lypcl} by putting  back the units $\lambda_{Q}=\omega_0\tilde{\lambda}_{Q}$. Then, taking $n_0=I/\hbar$, and inserting the definitions of $\omega_0$ and $\tilde{\epsilon}$ we have
$$
\lambda^3_{Q} = \frac{1}{2}\epsilon \alpha^{\frac{2-\gamma}{2}}m^{\frac{-2-\gamma}{2}}I^{\gamma-2},
$$
which is exactly the Lyapunov we get for the classical Hamiltonian in Eq.~\eqref{eq:Hclass} with $\tilde{q}(e^{i\Theta})=2\cos\Theta$.

\section{Numerical simulations of kicked systems}
\label{sec:numerical}

We now move to verify with numerical simulations all the theoretical results derived in the previous sections. In particular, working in the classical or quantum tangent space allows us to derive the asymptotic exponential growth of trajectories separations or the OTOCs, but it says nothing explicitly on the saturation of this divergence--- the prescrambling time. The saturation is expected when the angular separation reaches a value of order one, when a phase-space wavepacket would
cover the classical torus. Below we study numerical examples of the classical and quantum problems.

One way to realize the external white noise is to treat a kicked system
\begin{equation}
\eta(t)=\sum^{\infty}_{k=-\infty}r_k\delta(t-k\tau),
\label{eq:etaK}
\end{equation} 
with some kicking rate $\tau^{-1}$ and where $r_k$ are taken from a normal distribution of zero mean and variance $\tau\tilde{\epsilon}$. If the time between kicks is shorter than the unperturbed evolution and the Lyapunov time $\tau\ll\omega^{-1}_{n}\ll t_{\rm Lyp}$, then the external drive can be considered as a white noise. Stroboscopic drive with a constant magnitude corresponds to fundamental examples in the study of classical and quantum chaos. A well-known system is the Standard (Chirikov) map and its quantum equivalent--- the Quantum Kicked Rotor~\cite{Casati_etal,Chirikov_etal1988,Izrailev1990,Fishman_etal1982,Grempel_etal1982}.  

The case of random kicking is closely related to Chirikov `Typical Map'~\cite{ChirikovTypical,Frahm&Shepelyansky2009}, where the magnitudes of the kicks are given by a set of $T$ random variables which is repeated periodically. Frahm and Shepelyansky~\cite{Frahm&Shepelyansky2009} studied in detail the classical and quantum version of this  map. For the classical case they found a Lyapunov exponent that scales as $\tilde{\epsilon}^{1/3}$.

In what follows we treat the adimensional model
\begin{equation}
\tilde{H}(N,e^{i\Theta})=N^{\gamma}+2\eta(t)\cos\Theta,
\label{eq:Hnum}
\end{equation} 
where $\eta(t)$ is given in Eq.~\eqref{eq:etaK}. This randomly kicked system can be integrated numerically by applying the unitary operation $U_{\tau}(r)=e^{-iN^{\gamma} \tau}e^{-2ir\cos\Theta}$ between kicks, where $r$ is drawn from a normal distribution of zero mean and variance $\tilde{\epsilon}\tau$. We work in the eigenbasis of $N$, where operations of $\cos\Theta$ correspond to $2\langle n| \cos\Theta |n'\rangle= \delta_{n,n'+1}+ \delta_{n+1,n'}$ for $n,n'$ nonnegative. This can be done in Fourier space, as long as the system is far from the edges $n=0$ and $n=M$, with $M$ the size of the system. We verified that this is indeed a good approximation, by exact diagonalization of $\cos\Theta$. 

We consider the micro-canonical OTOC, where the system is initialized with $\ket{\psi_0}=\ket{n_0}$ and focus on the evolution of $C^2(t)=\bra{\psi_0}[\cos\Theta(t),N_0]^2\ket{\psi_0}$. For our localized initial wavefunction we have that 
\begin{equation}
C^2(t)=2\sum_n (n-n_0)^2 \left|\bra{n}U^{\dagger}\cos\Theta_0 U\ket{n_0}\right|^2,
\label{eq:Cnum}
\end{equation} 
with the evolution operator $U=\cdots U_{\tau}(r_3)U_{\tau}(r_2)U_{\tau}(r_1)$. 
In all the examples below we choose a kicking rate which is fixed with respect to the Lyapunov exponent, $\tau\sim 0.01 t_{\rm Lyp}(n_0,\tilde{\epsilon})$, to guarantee uncorrelated drive and allow a reasonable number of timesteps for observing a Lyapunov regime.

\subsection{Classical and quantum weakly, but randomly, kicked rotor}

The case of $\gamma=2$ (infinite potential well) can be considered as a rotor. The only difference is that for the latter, the angular momentum operator $N$ can assume negative values. In that case, one should modify the operation $\cos\Theta$ to account for negative values as well. We verified that the latter does not affect the results. 

\subsubsection{Classical}
As a reference, and demonstration of the theoretical description presented in Sec.~\ref{sec:classical}, we study the analog classical problem
\begin{equation}
H_{\cl}(I,\Theta)=\frac{I^2}{2}+2\eta(t)\tilde{\epsilon}^{1/2}\cos\Theta.
\end{equation}
The Hamilton equation yields the random map
\begin{eqnarray}
I_{t+dt} &=& I_{t}+2 r_t\sin\Theta_t,\\
\Theta_{t+dt} &=& \Theta_t+I_{t}dt \pmod {2\pi},
\end{eqnarray}
where $r_t$ is taken from a normal distribution of zero mean and variance $\tilde{\epsilon} dt$. We integrate this map for pairs of initial conditions, one initialized at $I_0=0$ and some random initial phase $\Theta_0=\alpha_0$ and the other is at a distance $\vec{u}_0$ from it. We fix the initial norm of $\vec{u}_0=10^{-8}$.  A pair of such initial conditions is integrated with the same realization of the noise.

\begin{figure*}
\centerline{\resizebox{0.45\textwidth}{!}{\includegraphics{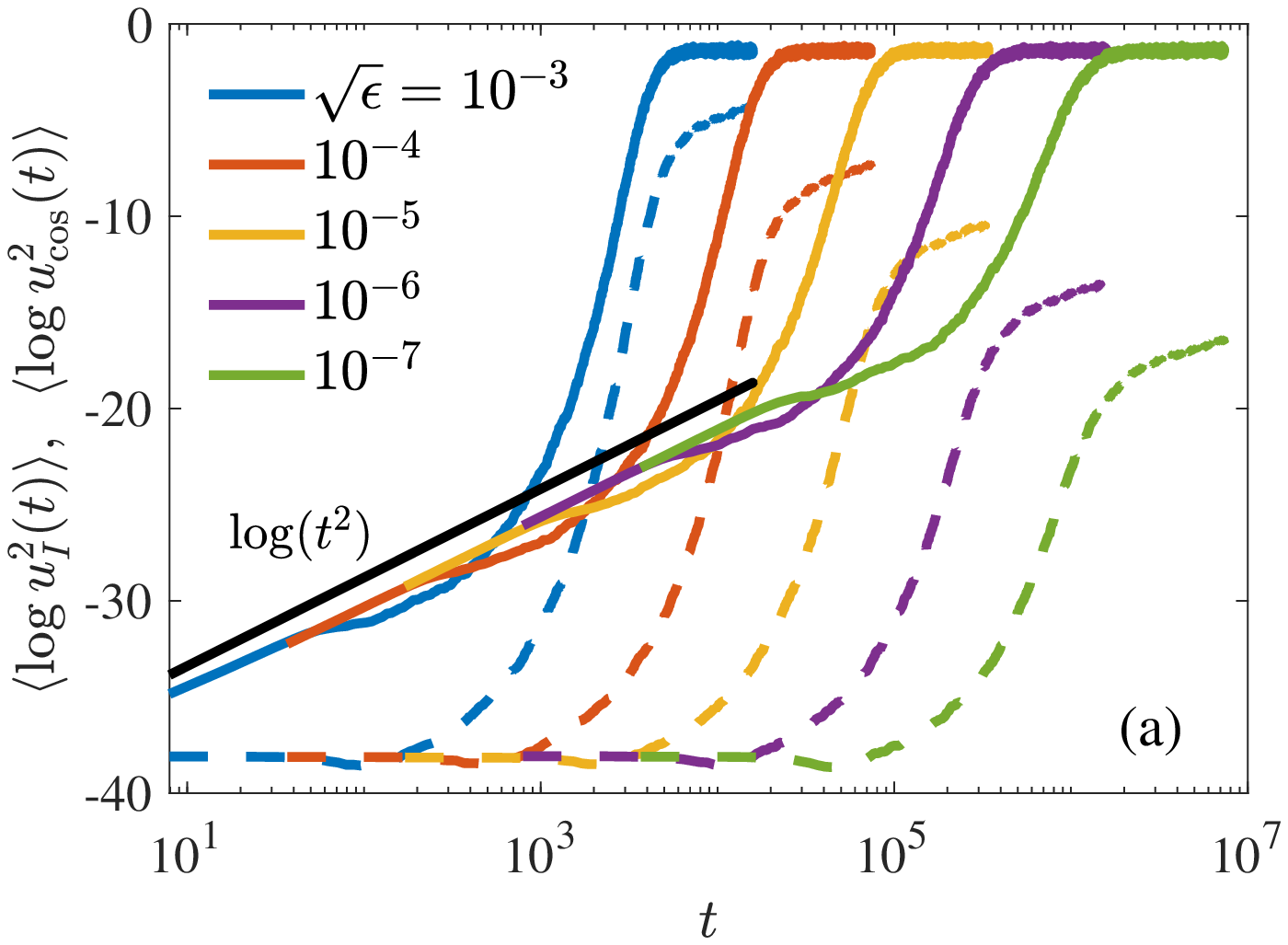}}
\hspace{1cm} 
\resizebox{0.45\textwidth}{!}{\includegraphics{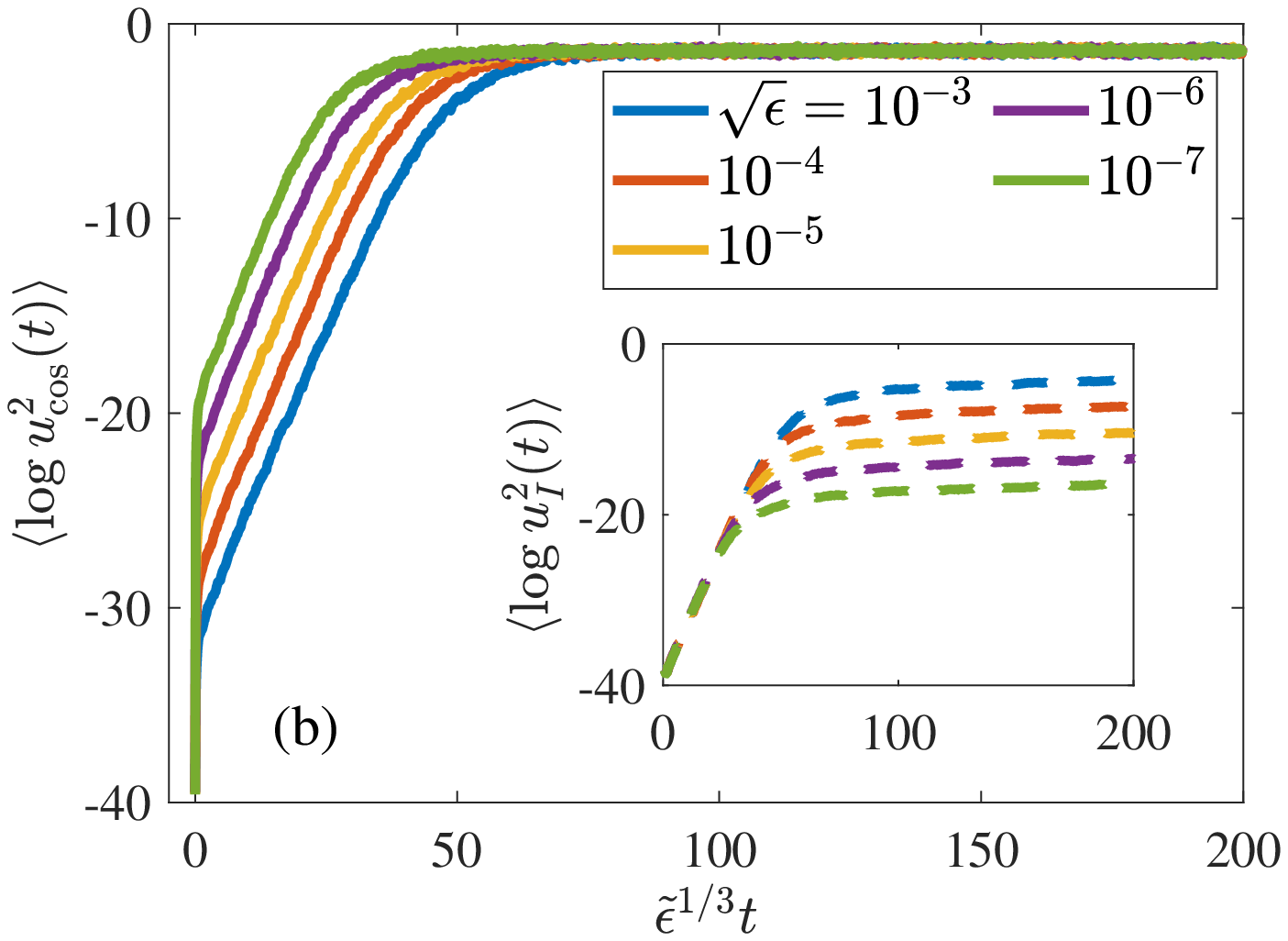}}}
\caption[]{ The separation of two initially close by trajectories for the {\em classical} randomly kicked rotor (averaged over 1000 pairs of trajectories). The solid lines corresponds to angular separation whereas the dashed lines indicate the difference in the action variables. Different curves correspond to different noise magnitude, $\tilde{\epsilon}^{1/2}$ outlined in the legend (smaller perturbations corresponds to longer saturation times). The initial separation is fixed.  In panel (a) the time axis is in logarithmic scale, whereas in panel (b) it is in linear scale and rescaled with $\tilde{\epsilon}^{1/3}t\sim \lambda_{\cl}t$.}
\label{fig:ClLyps}
\end{figure*}

In Fig.~\ref{fig:ClLyps}(a) we present separately the quenched evolution of the action separation $u_I$ (dashed curves) and the angular separation $u_{\cos}=\cos\Theta^{(1)}-\cos\Theta^{(2)}$ (solid curves). The figure shows all the three regimes discussed in Sec.~\ref{sec:classical}: at short times, the angular separation roughly grows in a linear fashion (ballistic regime), whereas the action separation changes little. At later times, an exponential growth starts in both coordinates and saturates when $u_{\theta}\sim O(1)$. By that time, still  $u_I\ll 1$, the more so the weaker the perturbation. In Fig.~\ref{fig:ClLyps}(b) we show, by collapsing the curves with rescaling time, that the rate of exponential growth is proportional to $\tilde{\epsilon}^{1/3}$, as expected.

\subsubsection{Quantum}

\begin{figure}
\centerline{\resizebox{0.45\textwidth}{!}{\includegraphics{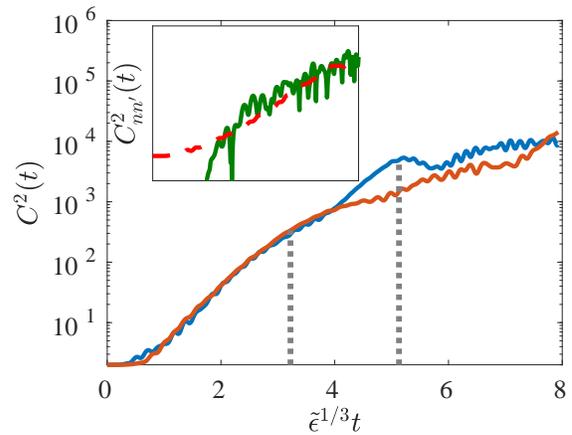}}}
\caption[]{ The OTOC growth for the quantum randomly kicked rotor, for two different initial conditions, $n_0=8191$ (blue) and $n_0=255$ (red), and fixed magnitude of the  perturbation $\tilde{\epsilon}^{1/2}=100$. The time is rescaled with $\tilde{\epsilon}^{1/3}$. For smaller $n_0$ the Ehrenfest time (indicated by the dotted grey lines), measured in Lyapunov times, is shorter. The inset shows off-diagonal components of the commutator $C_{n_0,n_0+10}$ in green solid line. The growth of the diagonal elements, shifted along the $y$-axis for reference, are shown in red dashed line ($n_0=255$).}
\label{fig:Lyps2}
\end{figure}

\begin{figure}
\centerline{\resizebox{0.45\textwidth}{!}{\includegraphics{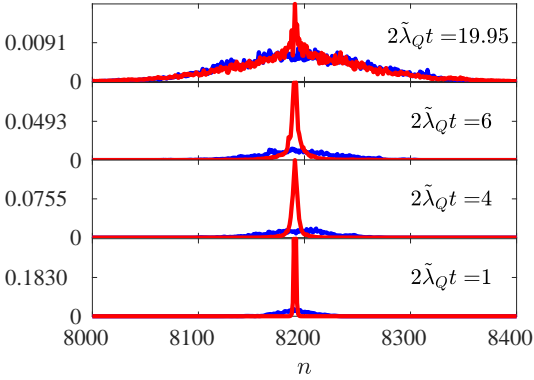}}}
\caption[]{The spreading of the initial wave-function $U(t)\ket{n_0}$ (blue) and $\bra{n}\cos\Theta(t)\ket{n_0}$ (red) that controls the OTOC growth (see Eq.~\eqref{eq:Cnum}). Initially, the latter is narrower than the former until they meet each other at later times. The initial eigenstate refers to $n_0=2^{13}-1$, where the size of the system is $M=2^{14}$.  The curves are averaged over 76 realizations of the noise.}
\label{fig:hist}
\end{figure}

Let us now move to the quantum problem. In Figure~\ref{fig:Lyps2} we show the evolution of the OTOC in Eq.~\eqref{eq:Cnum} for two different initial conditions with the same perturbation strength $\tilde{\epsilon}$. Times are rescaled with $\tilde{\epsilon}^{1/3}$. The curves show how the prescrambling time, measured in Lyapunov times, decreases with decreasing $n_0$.  As explained in the beginning of Sec.~\ref{sec:OTOC}, we shall have in mind an initial condition of a wavepacket around an eigenstate $\ket{n_0}$, rather than strictly the eigenstate. We have thus verified that--- starting with $\ket{\psi_0}=\ket{n_0}$, at later times the off-diagonal terms of the commutator square grow roughly as the diagonal terms (see inset of Fig.~\ref{fig:Lyps2}). Fig.~\ref{fig:hist} shows how energy diffuses little during the Lyapunov regime, this is equivalent to the small diffusion of the tori in the classical case.  It would be interesting to check that a wavepacket in the coherent-state or Wigner representation indeed fills the torus in an Ehrenfest time, and diffuses subsequently~\cite{Lam&Kurchan2014,Goldfriend&Kurchan2019}.

\begin{figure*}
\centerline{\resizebox{0.45\textwidth}{!}{\includegraphics{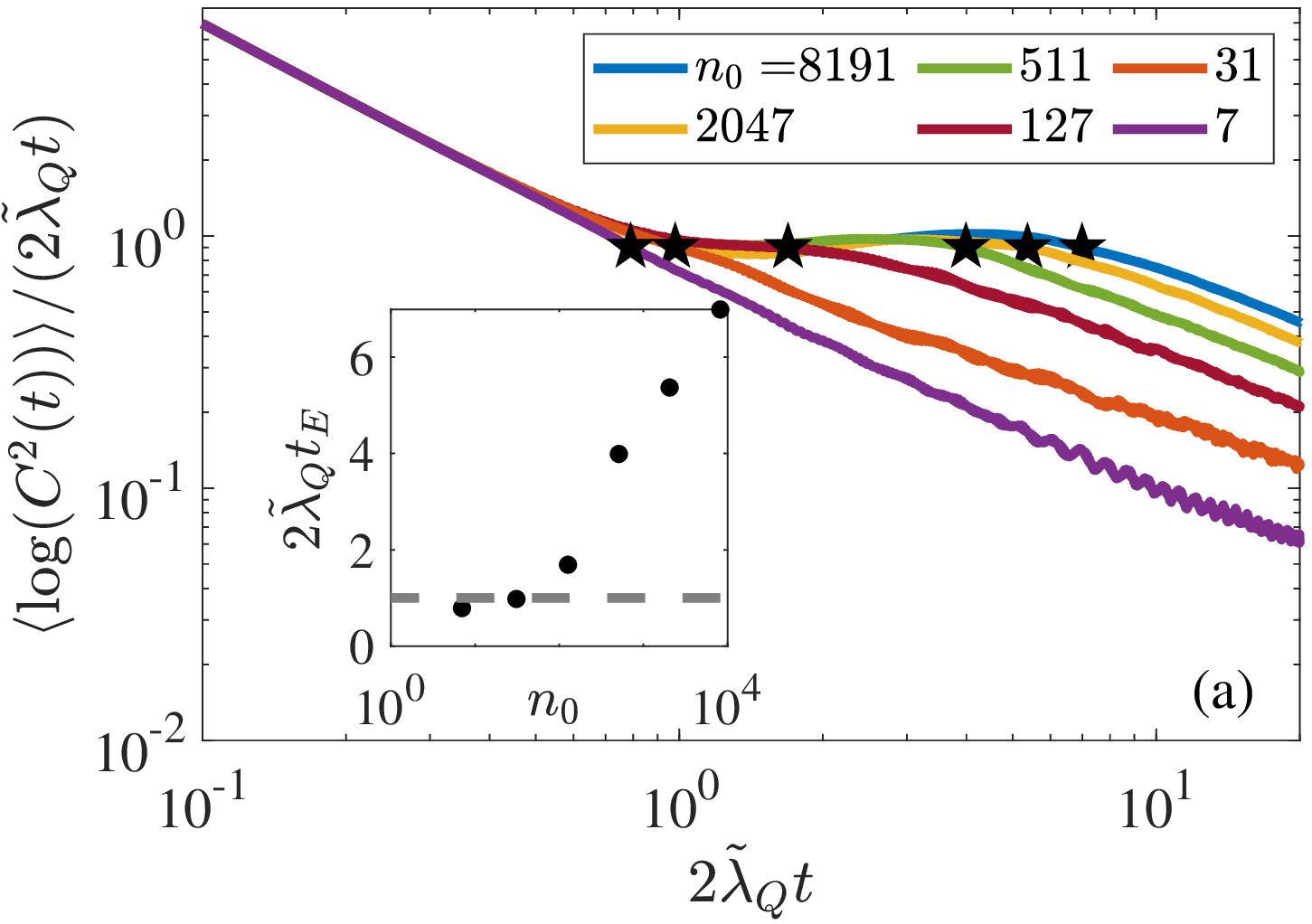}}
\hspace{1cm} 
\resizebox{0.45\textwidth}{!}{\includegraphics{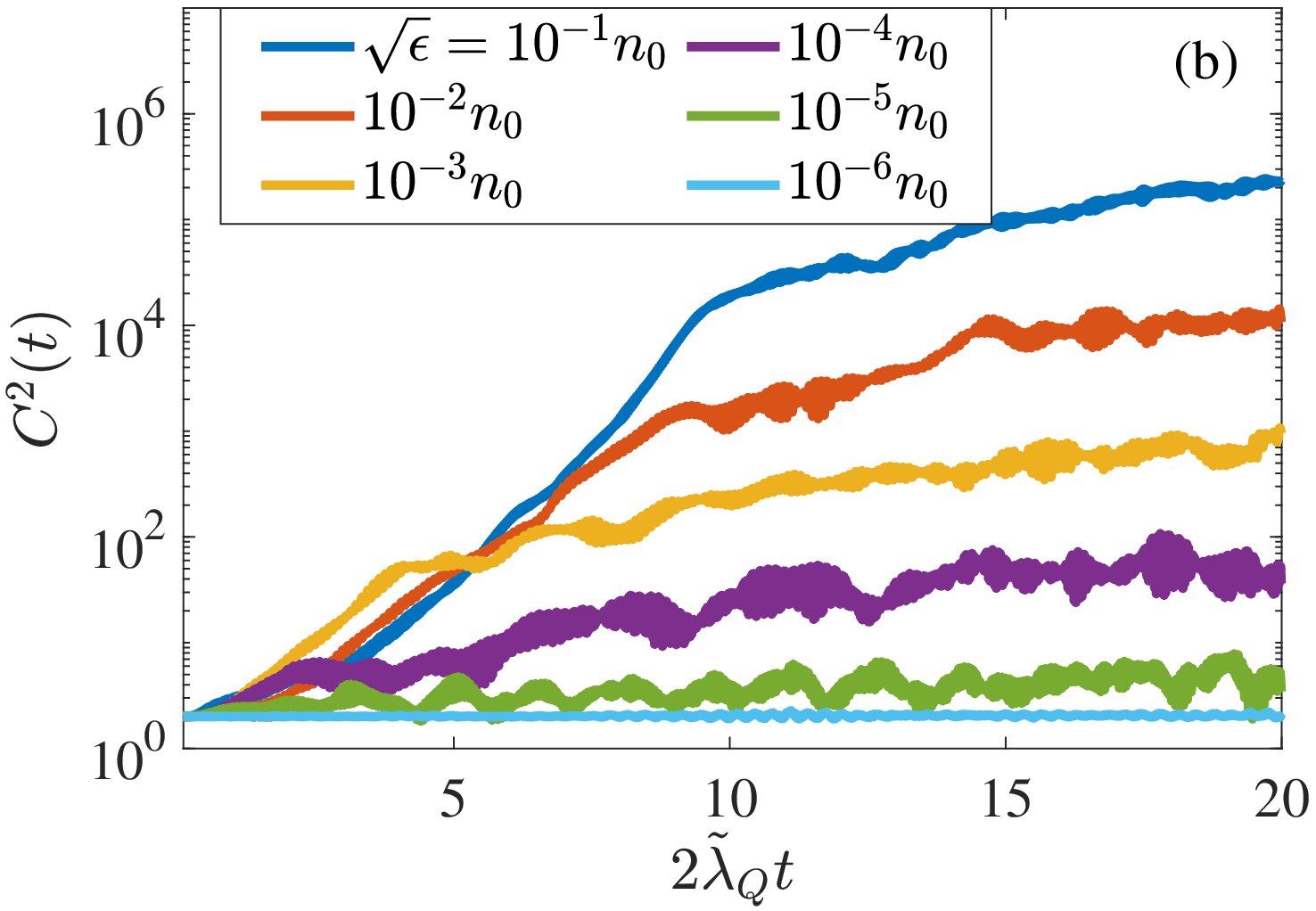}}}
\caption[]{(a) The (quenched) growth rate of the OTOC for a fixed relative perturbation strength $\tilde{\epsilon}^{1/2}/n_0$ as a function of time rescaled by the semiclassical Lyapunov exponent (averaged over 86 noise realizations). For larger initial energy level $n_0$ the Lyapunov regime is longer, below a certain $n_0$ the Lyapunov regime disappears. The inset shows how $\tilde{\lambda}_Qt_E$, with $t_E$ being the prescrambling time (stars in the main figure), increases with the logarithm scale of $n_0$, indicating a linear trend. (b) The growth of the OTOC (one noise realization) for different relative perturbation strength and fixed initial state $n_0=8191$ (smaller perturbation corresponds to lower saturation levels).}
\label{fig:Lyps2ab}
\end{figure*}

In Fig.~\ref{fig:Lyps2ab}(a) we show the growth of the OTOC for different initial energy levels $n_0$ and fixed relative perturbation $\epsilon^{1/2}=5\times 10^{-3}n_0$ (In this example we choose to keep $\tilde{\epsilon}/n^2_0$ fixed rather than $\tilde{\epsilon}$  since the former controls the smallness of the perturbation, as the energy is almost constant throughout the evolution). The OTOCs show an exponential growth with essentially  the classical Lyapunov exponent for a time window--- the Lyapunov regime---  that roughly starts at one Lyapunov time and ends at the prescrambling time $t_E$. In the inset of Fig.~\ref{fig:Lyps2ab}(a) we show that the latter is proportional to $\log n_0$ Lyapunov times. For $\log n_0 \sim 1$  there is no Lyapunov regime. This is the usual situation for large $\tilde \epsilon$. For smaller $\tilde \epsilon$ the Lyapunov time becomes large, and we must correct the initial size  $\ell_q(\tilde \epsilon)\sim \tilde{\epsilon}^{-1/3}$, according to the estimate above (see discussion in Sec.~\ref{sec:summ}). 

We check how  the prescrambling time depends on the perturbation strength. In Fig.~\ref{fig:Lyps}(b) we show the evolution of $C^2(t)$ for various magnitudes of external noise and fixed initial condition.
The Lyapunov regime gets shorter with decreasing $\tilde \epsilon$, and for small enough perturbation it vanishes. This behavior resembles the one observed  for the classical model in Fig.~\ref{fig:ClLyps}.

\subsection{Other randomly kicked integrable models}

\begin{figure*}
\centerline{\resizebox{0.45\textwidth}{!}{\includegraphics{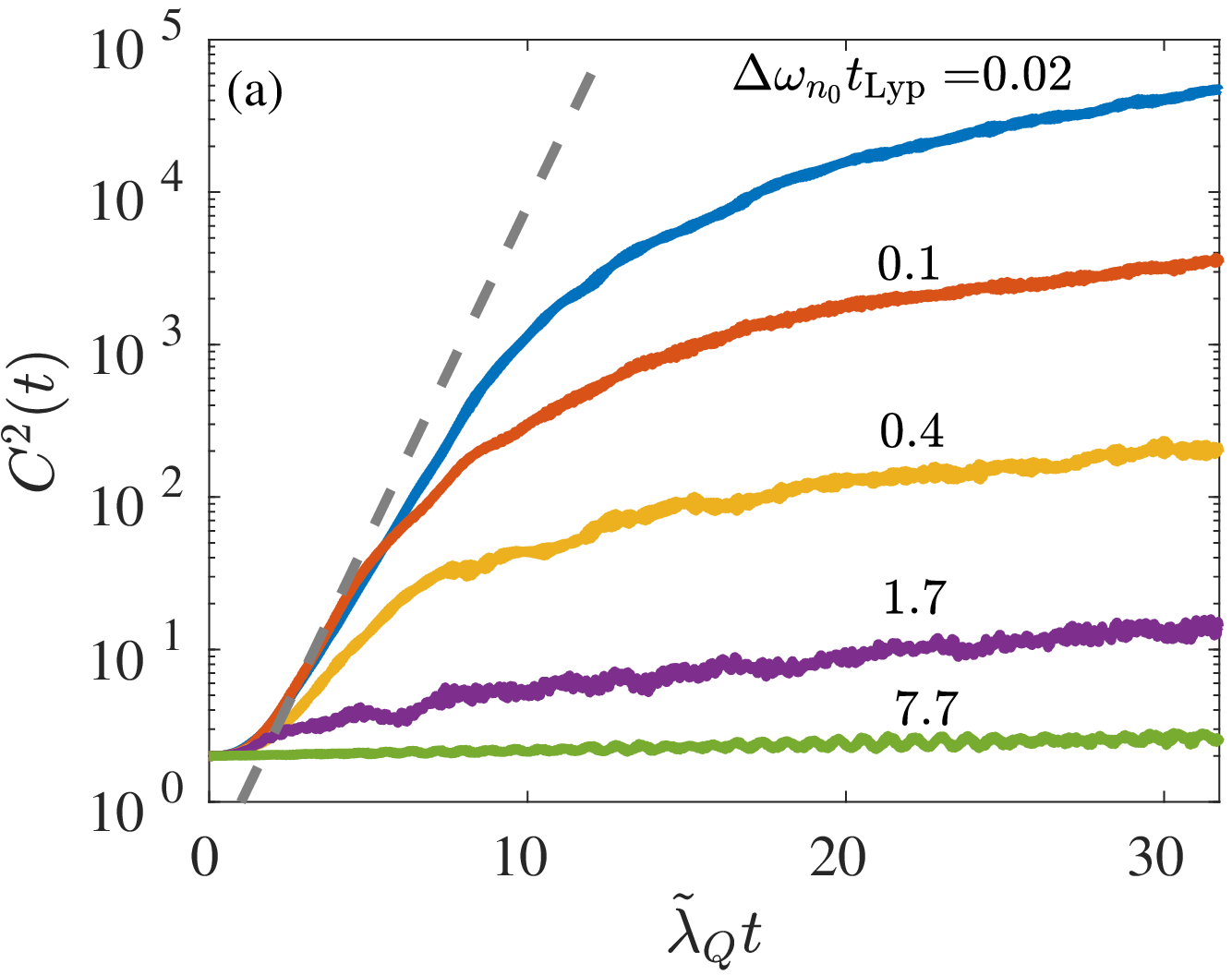}}
\hspace{1cm} 
\resizebox{0.45\textwidth}{!}{\includegraphics{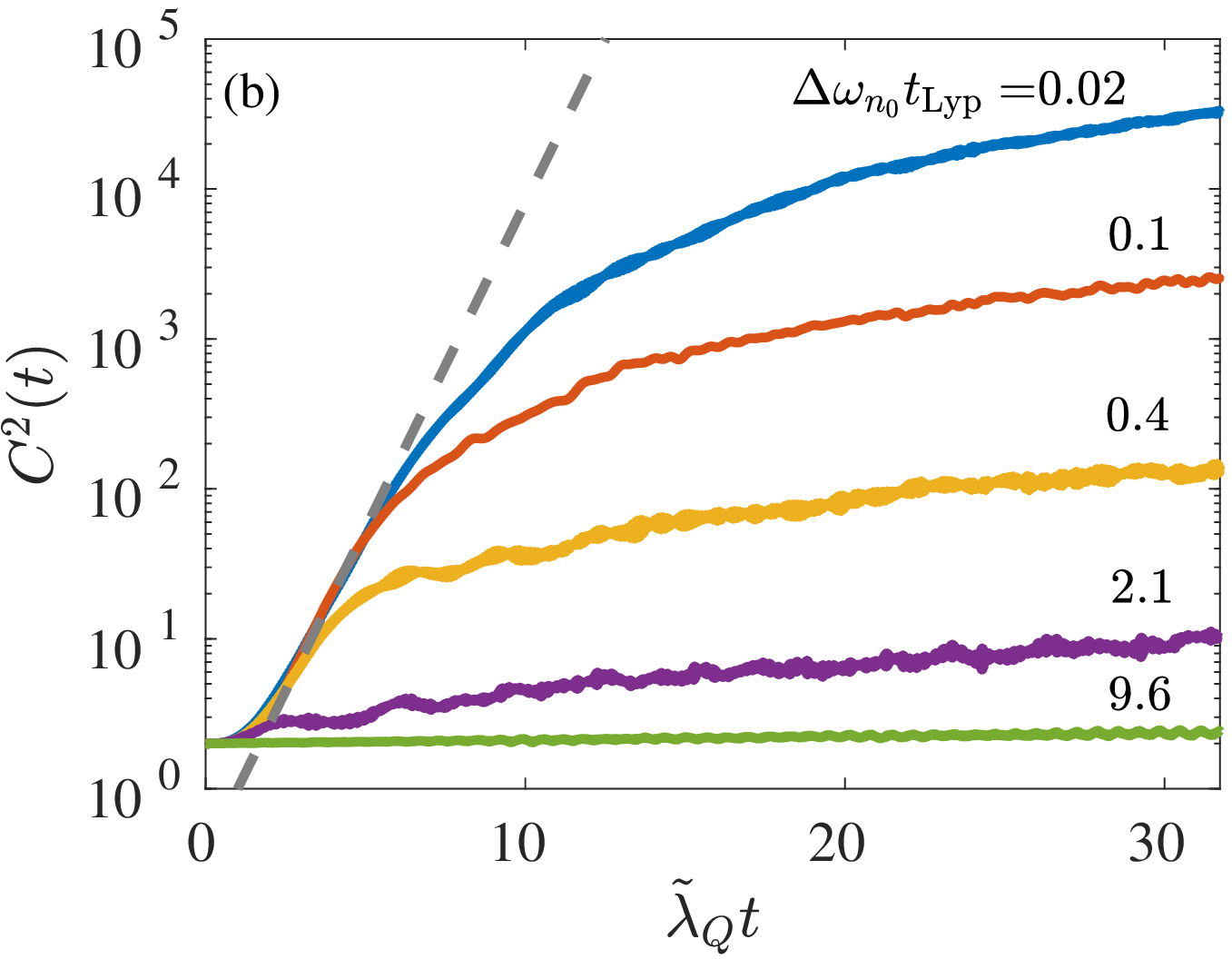}}}
\caption[]{ The (quenched) growth rate of the OTOC under the evolution of the Hamiltonian in Eq.~\eqref{eq:Hnum} with $\gamma=4/3$ and $\gamma=5/2$ in panels (a) and (b) respectively. The curves are averaged over 17 realizations of the random drive. The different curves corresponds to fixed initial condition $n_0=2^9-1$ and varying perturbation $\tilde{\epsilon}$. For each curve the time is rescaled with the Lyapunov exponent $\tilde{\lambda}_{Q}(\tilde{\epsilon})=2\tilde{\epsilon}^{1/3}\left[\gamma(\gamma-1)n^{\gamma-2}_0\right]^{2/3}$. The gray dashed line is $\exp(t-1)$ shows clearly the validity of the theoretical prediction of $\tilde{\lambda}_Q$. The values of $\tilde{\epsilon}$ for the different curves are $(10^{-1},10^{-2},10^{-3},10^{-4},10^{-5})\times n_0^{\gamma}$, where larger $\tilde{\epsilon}$ corresponds to longer (rescaled) saturation times.}
\label{fig:Lyps}
\end{figure*}

In Fig.~\ref{fig:Lyps} we present results for the case of $\gamma=4/3$ and $\gamma=3/2$ that correspond to the Bohr-Sommerfeld approximation of an integrable part with power-potential $q^4$ and $q^6$ respectively; see Eq.~\eqref{eq:Hclass} (the perturbation part is taken with $\mu=0$, as $N$ is roughly fixed during the Lyapunov regime).  Similar to the previous examples, the figure shows that the quantum Lyapunov exponent follows the classical one, $\tilde{\lambda}_Q\sim \tilde{\epsilon}^{1/3}n^{2(\gamma-2)/3}_0$. The exponential growth starts after $\sim 1$ Lyapunov time until it saturates at later time.  The figure also illustrates how the Lyapunov regime vanishes at
sufficiently weak  perturbation according to Eq.~\eqref{eq:Zcrit}.  The relevant quantity $\Delta\omega_{n_0}=\gamma(\gamma-1)n^{\gamma-2}_0$ is fixed by the initial condition, whereas the Lyapunov time increases with decreasing $\tilde{\epsilon}$. As the ratio $\Delta\omega_{n_0}/\tilde{\lambda}_Q$ increases the exponential growth saturates earlier. Once this ratio is $\sim O(1)$ we do not observe an exponential growth.
The ballistic regime suffices to scramble over the torus, as explained above.

{\color{black} 
\section{The quantum bound on chaos}
\label{sec:bound}

In the current Paper we have focused on a microcanonical version of the OTOC, namely, the expectation value of $\bra{n_0}(C^{\Theta})^2\ket{n_0}$. Recently, it was shown that the quantum Lyapunov exponent (defined by the growth rate of the OTOC) is bounded in {\em thermal} systems as $\beta\hbar\lambda_{\mbox{\tiny  T}}\leq 2\pi$~\cite{Maldacena_etal2016}. 
We shall argue that, at least in our model, the quantum limitation to chaos is imposed by blocking one by one the Lyapunov regimes of the degrees of freedom
that would yield the largest Lyapunov divergencies.

Let us start by performing a canonical averaging:
\begin{equation}
\lambda_{\mbox{\tiny  T}} = \frac{1}{t} \ln {\mbox{Tr}} \left\{ [A(t),B_0]^2 \; e^{-\beta \omega_0\hbar \;  \tilde H_{\rm int}(N)} \right\}/Z,
\label{eq:annLyp}
\end{equation}
\begin{equation}
n_{\mbox{\tiny  T}}(\beta \omega_0\hbar)= {\mbox{Tr}} \left\{ N  \; e^{-\beta \omega_0\hbar \; \tilde H_{\rm int}(N)} \right\}/Z,
\label{eq:nbar}
\end{equation}
where we put back the energy and time scales, $\hbar\omega_0$ and $\omega_0^{-1}$ respectively. The long time limit in the annealed averaging of Eq.~\eqref{eq:annLyp} has to be taken with care, or alternatively, one can make a `quenched' calculation by taking the expectation of the {\em logarithm} of the squared commutator. The canonical averaging in Eq.~\eqref{eq:nbar} imposes a relation $\beta\hbar\omega_0\equiv k(n_{\mbox{\tiny  T}})$, which is a decreasing function of $n_{\mbox{\tiny  T}}$. Hence, if we evaluate the averaging in Eq.~\eqref{eq:annLyp} with $n_{\mbox{\tiny  T}}$ we obtain:
\begin{equation}
 \beta\hbar\lambda_{\mbox{\tiny  T}} = k(n_{\mbox{\tiny  T}})\tilde{\lambda}_Q(n_{\mbox{\tiny  T}},\tilde{\epsilon})
 \equiv g(n_{\mbox{\tiny  T}},\tilde{\epsilon}).
\label{adimlyap2}
\end{equation}

The fact that there should be at all a Lyapunov regime at a given finite value of $n_0$ already implies that the adimensional quantity $\beta\hbar\lambda_{\mbox{\tiny T}}$  scales as a finite number, and the system is a `rather good scrambler'  between Lyapunov and prescrambling times. We now derive a general semiclassical expression for the {\em adimensional} function $g(n_{\mbox{\tiny  T}},\tilde{\epsilon})$ and show that it grows, for a given $\tilde \epsilon$, as $n_{\mbox{\tiny  T}}$ decreases: the quantization of $n_{\mbox{\tiny  T}}$ will provide a bound. {The  system is then a relatively good scrambler, in the sense that $   \beta\hbar\lambda_{\mbox{\tiny  T}}$ reaches, at low $T$, a {\em finite} (albeit small) $\tilde \epsilon$-dependent value}. Note that although $\tilde\epsilon$ depends on $\hbar$ and $\omega_0$, we can change $n_{\mbox{\tiny  T}}$ independently by varying the temperature.


For a general Hamiltonian in Eq.~\eqref{eq:HQ}, the scalings for the adimensional Lyapunov and the assumption of weak perturbation are given respectively according Eq.~\eqref{eq:Lypcl} and Eq.~\eqref{eq:weak}, replacing $I_0\rightarrow n_0$. 
The adimensional relation in Eq.~\eqref{adimlyap2} then reads
\begin{equation}
\beta \hbar\lambda_{\mbox{\tiny T}}  =  \frac{\tilde{\lambda}(n_{\mbox{\tiny  T}},\tilde\epsilon)}{k^{-1}(n_{\mbox{\tiny  T}})}=\left(\frac{\tilde{\epsilon}\bar{q}^2(n_{\mbox{\tiny  T}})}{\tilde{H}''_{\rm int}(n_{\mbox{\tiny  T}})n_{\mbox{\tiny  T}}^3}\right)^{1/3}\left(\frac{n_{\mbox{\tiny  T}}\tilde{H}''_{\rm int}(n_{\mbox{\tiny  T}})}{k^{-1}(n_{\mbox{\tiny  T}})}\right).
\end{equation}
Now, the first brackets cannot be too large according to Eq.~\eqref{eq:weak}, and, if we do not want to violate the bound~\cite{Maldacena_etal2016}, then the second brackets must be a decreasing function of $n_{\mbox{\tiny  T}}$. Hence, the most chaotic system corresponds for $n_{\mbox{\tiny  T}}=O(1)$, at which the Lyapunov regime also vanishes. We can verify this explicitly for the BS of $\tilde{H}_{\rm int}$, for which we find $\beta \hbar\lambda_{\mbox{\tiny T}}=\tilde{\epsilon}^{1/3}k(n_{\mbox{\tiny  T}})n_{\mbox{\tiny  T}}^{(\gamma-2)/3}$ with $0<\gamma\leq 2$.  We know, however, that when $n_0$ becomes of $O(1)$, the Lyapunov regime shrinks to zero.

 In a many-body system, the mechanism for the quantum bound may be hence understood as follows:
consider a system consisting  of $M$  copies of our {\em integrable} model, having values of $\omega_0= \omega^{(1)}_0> ... >\omega^{(M)}_0\sim0$, with $\omega^{(i)}_0$ spanning an interval that goes down to zero. The system is at temperature $T$, so that the corresponding average quantum numbers are $ n^{(1)}_{\mbox{\tiny  T}}< ...<n^{(M)}_{\mbox{\tiny  T}}$. The coupling introduces perturbations with $\tilde{\epsilon}^{(i)}$. 
Importantly, the global Lyapunov exponent is dominated by the largest of individual ones.

 Consider then  choosing the adimensional $\tilde \epsilon^{(i)}= \tilde \epsilon^{(1)}$, the same $\forall i$. 
At each  temperature some subsystems will have $ n^{(i)}_{\mbox{\tiny  T}}<1$, and will thus not contribute with a Lyapunov regime. Hence, the global Lyapunov exponent corresponds to the one of the systems $n^*_{\mbox{\tiny  T}} \sim 1$ that is just about to lose its Lyapunov regime by quantum effects, which one depends on the value of $T$. 
This in turn means that, as $T \rightarrow 0$, the combination $\beta \hbar \lambda_T$ remains a number of $O(1)$, albeit small (because $\tilde \epsilon $ is small).

\section{Conclusion and outlook}
\label{sec:dis}

In the current manuscript we have focused on the case of a one degree of freedom integrable model, which is weakly driven by an external white noise.  For the classical counterpart~\cite{Lam&Kurchan2014}, we know that the mechanism of the exponential growth `over the torus' and saturation is in fact generic, and is also relevant in higher dimensions. The same is expected for the quantum problem and the growth of the OTOC. 
The formalism of the quantum tangent space lays the groundwork to study such generalization.

Our main  motivation is to understand the properties of an isolated quasi-integrable models by mimicking the many-body integrability breaking coupling by some external noise. However, the model we study concerns only classical noise acting on a quantum system, which might not be suitable to capture all the effects of quantum couplings. One, rather primitive, way to account for noise with quantum origins is to consider correlated instead of white noise. This can be addressed theoretically, as was done for the classical problem~\cite{Lam&Kurchan2014}, or numerically--- by reducing the kicking rate with respect to the unperturbed evolution and the Lyapunov time. A more serious way to take  into account the  quantum origins of noise is to start with a 1-dimensional model which is coupled to ensemble of linear oscillators and employ the Feynman-Vernon, Caldeira Leggett method, e.g., as in Ref.~\cite{Tuziemski2019}.  Within this formalizm, the assumption of Markovianity shall lead to  a Lindblad-like operator, which accounts for the coupling to the bath~\cite{Brasil2013}. We expect this term to enter within the tangent space formalizm, however, we note that such a new term should not modify the results, as it can be derived from a classical noise on a quantum system~\cite{Rowlands&Lamacraft2018}.

Our results highlight the meaning of the Ehrenfest time as the time at which the wave character of a quantum system plays an important role. We have found that  for the case of very weak perturbations and that of  very small quantum numbers, the mechanism for exponential growth is turned off by quantum effects that originate in the discreteness of the spectrum, ultimately the uncertainty principle. This phenomenon is absent in the classical case, where two initial conditions can be arbitrarily close to each other. Our results show how the energy (or any other quasi-constant of motion) 
does not diffuse significantly during the Lyapunov regime (Fig.~\ref{fig:hist}).  We have not explicitly
studied the scrambling of the $\cos\Theta$ operator, i.e. an analogous quantum picture for the solid curves in Fig.~\ref{fig:ClLyps}. One interesting future direction is thus to see how a phase-space 
wavepacket  spreads throughout the torus, when described in the coherent-state representation.  

The formalism we derived--- the quantum tangent space, e.g., Eq.~\eqref{eq:ODEC}--- might be useful to address operator growth in other set-ups. An interesting generalization might be to consider a quantum chain of bosons, where the tangent space is written in terms of the ladder operators $a^{(\alpha)}$, replacing $e^{-i\Theta}$, and occupation numbers $N^{(\alpha)}=(a^{(\alpha)})^{\dagger}a^{(\alpha)}$. For such systems, apart of the exponential growth rate, one can also consider the butterfly velocity, i.e., the rate at which $[N^{(\alpha)},N_0^{(\beta)}]^2$ depends on time as a function of sites $\alpha$ and $\beta$.

Finally, it will be interesting to test the implications of our model against isolated quasi-integrable systems. In particular the scaling of the Lyapunov exponent with the effective perturbation strength, and its comparison to the thermalization time of the system.  
The latter has been recently measured in a cold-atom system with a tunbable integrability breaking interactions~\cite{Tang_etal2018}.

~\\~

\begin{acknowledgments}
We thank  Laura Foini,  Eduardo Fradkin, Andrea Gambassi, Stefano Ruffo, Dima Shepelyansky, Alessandro Silva, and Denis Ullmo for helpful discussions and suggestions, and Doron Cohen for pointing out relevant references for this work. TG and JK are supported by the Simons Foundation Grant No. 454943.
\end{acknowledgments}


\onecolumngrid
\appendix

\section{Dynamics in the space of Poisson Brackets}
\label{app:classical}

Below we derive the equations which govern the dynamics in the space of Poisson brackets. This dynamics is equivalent to the one of the tangent space. We treat the case of canonical and non-canonical variables   

\subsection{Canonical variables}

In relation to the problem studied in the paper, we consider action-angle variables $(I,\Theta)$. The derivations holds for any canonical variables, e.g., coordinates and momentum $(q,p)$, and for many-degrees of freedom. 

Since the Poisson brackets act as derivatives
\begin{equation}
\frac{\partial}{\partial I}=-\{\cdot,\Theta\}, \,\,\, 
\frac{\partial}{\partial \Theta}=\{\cdot,I\},
\end{equation}
they also have a corresponding chain rule: for a function $F(I,\Theta)$
\begin{equation}
\{F,\Theta_0\}=-\frac{\partial F}{\partial I_0}=-\frac{\partial I}{\partial I_0}\frac{\partial F}{\partial I}-\frac{\partial \Theta}{\partial I_0}\frac{\partial F}{\partial \Theta}=-\{I,\Theta_0\}\{F,\Theta\}+\{\Theta,\Theta_0\}\{F,I\}
\label{eq:chain1}
\end{equation}
\begin{equation}
\{F,I_0\}=\frac{\partial F}{\partial \Theta_0}=\frac{\partial I}{\partial \Theta_0}\frac{\partial F}{\partial I}+\frac{\partial \Theta}{\partial \Theta_0}\frac{\partial F}{\partial \Theta}=-\{I,I_0\}\{F,\Theta\}+\{\Theta,I_0\}\{F,I\}
\label{eq:chain2}
\end{equation}
where the subscript $0$ refers to values at initial time. In the last equality we use the fact that the Poisson brackets are canonically invariant--- taking them with respect to the canonical variables $(\Theta_0,I_0)$ or $(\Theta,I)$ is the same.

From the above relations we find
\begin{multline}
\frac{d}{dt}
\begin{pmatrix}
\{I,\Theta_0\}\\
\{\Theta,\Theta_0\}\\
\{I,I_0\}	\\
\{\Theta,I_0\}	
\end{pmatrix}=
\begin{pmatrix}
\{\{I,H\},\Theta_0\}\\
\{\{\Theta,H\},\Theta_0\}\\
\{\{I,H\},I_0\}	\\
\{\{\Theta,H\},I_0\}	
\end{pmatrix}=\\
=
\begin{pmatrix}
-\{\{I,H\},\Theta\}&\{\{I,H\},I\}&0 &0\\
-\{\{\Theta,H\},\Theta\}&\{\{\Theta,H\},I\}&0 &0\\
0&0&-\{\{I,H\},\Theta\}&\{\{I,H\},I\}	\\
0&0&-\{\{\Theta,H\},\Theta\}& \{\{\Theta,H\},I\}	
\end{pmatrix}\begin{pmatrix}
\{I,\Theta_0\}\\
\{\Theta,\Theta_0\}\\
\{I,I_0\}	\\
\{\Theta,I_0\}	
\end{pmatrix}.
\end{multline}
Since the upper and lower blocks are identical, and since the initial condition is 
$$
(\{I_0,\Theta_0\},\{\Theta_0,\Theta_0\},\{I_0,I_0\},\{\Theta_0,I_0\})=(-1,0,0,1),
$$
it is sufficient to consider only the first two entries
\begin{equation}
\frac{d}{dt}
\begin{pmatrix}
\{I,\Theta_0\}\\
\{\Theta,\Theta_0\}
\end{pmatrix}=
\begin{pmatrix}
\{\{H,I\},\Theta\}&-\{\{H,I\},I\}\\
\{\{H,\Theta\},\Theta\}&-\{\{H,\Theta\},I\}	
\end{pmatrix}\begin{pmatrix}
\{I,\Theta_0\}\\
\{\Theta,\Theta_0\}	
\end{pmatrix}.
\label{eq:var2}
\end{equation}

\subsection{Non-canonical variables}

We now consider the case when the pair of variables are not canonically conjugate. Instead of working in the action-angle space $(I,\Theta)$, we change coordinates to $(I,g(\Theta))$. Then, the Poisson brackets are related to the derivatives according to 
\begin{equation}
-\{\cdot,\Theta\}=\frac{\partial}{\partial I}=-\frac{1}{g'}\{\cdot,g\},
\end{equation}
\begin{equation}
\frac{\partial}{\partial g}=\frac{1}{g'}\{\cdot,I\},
\end{equation}
where $g'\equiv \partial g(\Theta)/\partial\Theta$.
The chain rule relations are then 
\begin{equation}
\{F,g_0\}=\frac{1}{g'}\left(-\{I,g_0\}\{F,g\}+\{g,g_0\}\{F,I\}\right)
\end{equation}
\begin{equation}
\{F,I_0\}=\frac{1}{g'}\left(-\{I,I_0\}\{F,g\}+\{g,I_0\}\{F,I\}\right).
\end{equation}
In analogy to Eq.~\eqref{eq:var2}, we have now
\begin{equation}
\frac{d}{dt}
\begin{pmatrix}
\{I,g_0\}\\
\{g,g_0\}
\end{pmatrix}=
\frac{1}{g'}
\begin{pmatrix}
\{\{H,I\},g\}&-\{\{H,I\},I\}\\
\{\{H,g\},g\}&-\{\{H,g\},I\}	
\end{pmatrix}\begin{pmatrix}
\{I,g_0\}\\
\{g,g_0\}	
\end{pmatrix}.
\label{eq:vargI}
\end{equation}
We note that  
\begin{equation}
\{\{H,I\},g\}=-\{\{g,H\},I\}-\{\{I,g\},H\}=\{\{H,g\},I\}+\dot{g}',
\label{eq:gprimedot}
\end{equation}
that is, the matrix includes full-time-derivatives of $g'$. The matrix appearing in Eq.~\eqref{eq:vargI} cannot imply symplectic dynamics, as the transformation $(I,\Theta)\rightarrow(I,g)$ is not a canonical one. We can get a symplectic dynamics by considering the vector 
$$
\begin{pmatrix}
\{I,g_0\}\\
\{g,g_0\}/g'
\end{pmatrix}, \text{ which satisfies the relation }
\begin{pmatrix}
\{I,g_0\}\\
\{g,g_0\}/g'
\end{pmatrix}=g'_0 \begin{pmatrix}
\{I,\Theta_0\}\\
\{\Theta,\Theta_0\}
\end{pmatrix}.
$$
The time-derivative of this new vector is identical to the one in Eq.~\eqref{eq:var2}, and can be written as 
\begin{equation}
\frac{d}{dt}
\begin{pmatrix}
\{I,g_0\}\\
\{g,g_0\}/g'
\end{pmatrix}=
\frac{1}{g'}
\begin{pmatrix}
\{\{H,I\},g\}&-\{\{H,I\},I\}g'\\
\{\{H,g\},g\}/g'&-\{\{H,I\},g\}	
\end{pmatrix}
\begin{pmatrix}
\{I,g_0\}\\
\{g,g_0\}/g'	
\end{pmatrix},
\label{eq:varIgg}
\end{equation}
where we use the relation in Eq.~\eqref{eq:gprimedot}.

\section{Chain-rule for commutators}

In the current Appendix we consider a general statement for a chain rule for commutators and its implication on the operators evaluated in the eigenbasis of $N$

\subsection{General relation}
\label{app:chain}

We prove the following general statement: for an analytic function (at some domain) $g$, and operators $A$ and $B$ we have
\begin{equation}
[A,g(B)]=\lim_{s\to 0}\frac{g(B+s[A,B])-g(B)}{s}.
\label{eq:ChainRule}
\end{equation}
Proof: for an integer power $g(x)=x^k$ we have the known formula (readily proven by induction) $[A,B^k]=\sum^k_{r=1} B^{r-1}[A,B]B^{k-r}$, which is equivalent to the expression in Eq.~\eqref{eq:ChainRule}:
\begin{equation}
\sum^k_{r=1} B^{r-1}[A,B]B^{k-r}=\lim_{s\to 0}\frac{(B+s[A,B])^k-B^k}{s}.
\end{equation}
The general result now follows, since $g$ is analytic then we have $g(B)=\sum a_k (B-x_0)^k$, and
$$
[A,g(B)]=\sum a_k [A,(B-x_0)^k]=\sum a_k \lim_{s\to 0}\frac{(B+s[A,B]-x_0)^k-(B-x_0)^k}{s}=
$$
$$
=\lim_{s\to 0}\frac{\sum a_k(B+s[A,B]-x_0)^k-\sum a_k(B-x_0)^k}{s}=\lim_{s\to 0}\frac{g(B+s[A,B])-g(B)}{s}.
$$
Eq.~\eqref{eq:ChainRule} induces a linear relation between $[A,g(B)]$ and $[A,B]$.

\subsection{Algebraic relations in the eigenbasis of $N$}
\label{app:eigenbasisN}

The basic relations we have are $e^{im\Theta}\ket{n}=\ket{n+m}$ and $\left[N,e^{im\Theta}\right]= m e^{im\Theta}$, that is, in the eigenbasis of $N$ we can write $\left(e^{i\Theta}\right)_{n,n'}=\delta_{n,n'+1}$ and $N_{n,n'}=n\delta_{n,n'}$. We use the chain-rule for commutators above  to calculate commutation with $N^{\gamma}$. Below we prove the relations:
\begin{eqnarray}
\label{eq:relation1}
\left(\left[\cdot,N^{\gamma}\right]\right)_{n n'}&=&\frac{n^{\gamma}-n'^{\gamma}}{n-n'}[\cdot,N]_{nn'},\\
\left(\left[e^{i\Theta},N^{\gamma}\right]\right)_{n n'}&=&\left((n-1)^{\gamma}-n^{\gamma}\right)\delta_{n,n'+1}.
\label{eq:relation2}
\end{eqnarray}
Proof: Since we know that in the action space $(N)_{n n'}=n\delta_{n n'}$ we can employ first order perturbation theory to write
$(N+s[\cdot,N])=U^{-1} D U $, where
$$
D_{n n}=n+s[\cdot,N]_{nn}, \,\,\, U_{n n'}=1+s\frac{[\cdot,N]_{nn'}}{n-n'}.
$$
Note that the last term does not diverge as $[\cdot,N]_{nn}=0$ since $N$ is diagonal. Therefore, to leading order in $s$ we have 
\begin{equation}
(N+s[\cdot,N])_{nn'}=n\delta_{nn'}+s\left(n\frac{[\cdot,N]_{nn'}}{n-n'}-n'\frac{[\cdot,N]_{nn'}}{n-n'}\right),
\end{equation}
and subsequently 
\begin{equation}
(N+s[\cdot,N])^{\gamma}_{nn'}= n^{\gamma}\delta_{nn'}+s\left(n^{\gamma}\frac{[\cdot,N]_{nn'}}{n-n'}-n'^{\gamma}\frac{[\cdot,N]_{nn'}}{n-n'}\right),
\end{equation}
which gives Eq.~\eqref{eq:relation1}
\begin{equation}
\left([\cdot,N^{\gamma}]\right)_{n n'}=\frac{n^{\gamma}-n'^{\gamma}}{n-n'}[\cdot,N]_{nn'}.
\end{equation}
Finally, the representation of $e^{i\Theta}$ in the eigenbasis of $N$ gives the relation in Eq.~\eqref{eq:relation2} 
\begin{equation}
\left(\left[e^{i\Theta},N^{\gamma}\right]\right)_{n n'}=\left((n-1)^{\gamma}-n^{\gamma}\right)\delta_{n,n'+1}.
\end{equation}

\section{Quantization of the power potential}
\label{app:quantization}

\subsection{Quantization of the integrable part}

We look at the general classical Hamiltonian 
\begin{equation}
H_{\cl, \ing}=\frac{p^2}{2m}+\alpha q^\nu.
\end{equation}
The action variable of this Hamiltonian can be calculated explicitly~\cite{Carinena_etal1993}:
\begin{equation}
I(H_{\cl, \ing})=s(\nu)\alpha^{-1/\nu}\sqrt{m}H^{\frac{2+\nu}{2\nu}}_{\cl, \ing},
\end{equation} 
where $s(\nu)=\sqrt{8\pi}\frac{\Gamma(1/\nu+1)}{\Gamma(1/\nu+3/2)}$ with the $\Gamma$ Euler function. This gives
\begin{equation}
H_{\cl, \ing}(I)= s^{\gamma}(\nu)\alpha^{1-\gamma/2}m^{-\gamma/2} I^{\gamma}\equiv K(m,\alpha,\nu) I^{\gamma},
\label{eq:HclI}
\end{equation}
with $\gamma\equiv\frac{2\nu}{2+\nu}$. Quantization of the classical Hamiltonian can be obtained by a rescaling procedure: we substitute $q\rightarrow (\hbar/b) q$, $p\rightarrow b p$, and we require that $H_{\cl, \ing}=f(\alpha,\nu,m,\hbar)(p^2+q^{\nu})$. One finds the rescaling parameter
\begin{equation}
b=\left(m\alpha\hbar^{\nu}\right)^{\frac{1}{2+\nu}}=\left(m\alpha\right)^{\frac{2-\gamma}{4}} \hbar^{\frac{\gamma}{2}},
\label{eq:b}
\end{equation}
and accordingly we can write $f(\alpha,\nu,m,\hbar)\equiv \hbar\omega_0\tilde{H}(N)$ with
\begin{equation}
\omega_0\equiv \alpha^{1-\frac{\gamma}{2}}m^{-\frac{\gamma}{2}}\hbar^{\gamma-1},
\label{eq:omg0}
\end{equation}
having dimensions of $\text{time}^{-1}$.
Therefore, a quantization of the integrable Hamiltonian is simply
\begin{equation}
H=\hbar\omega_0\tilde{H}(N).
\end{equation}

In the semi-classical limit, according to the Bohr-Sommerfeld quantization we shall substitute $I=\hbar N$ in Eq.~\eqref{eq:HclI}, which gives $\tilde{H}(N)=N^{\gamma}$. Finally, let us note that since $0<\nu<\infty$ we have that $0<\gamma<2$. For Harmonic oscillator we have $\nu=2$, $\gamma=1$, and for the infinite potential well $\nu\rightarrow\infty$, $\gamma=2$.

\subsection{Quantization of the perturbation part}

For the stochastic perturbation of the Hamiltonian we assume the classical form $\epsilon^{1/2} q \eta(t)$, where $\eta(t)$ has units of $\text{time}^{-1}$. Therefore, the dimensions of $\epsilon^{1/2}$ are $\text{energy}\cdot\text{time}^{1/2}\cdot\text{length}^{-1}$. Inserting the rescaling parameter $b$ for the coordinate variable, we have 
\begin{equation}
H_{\ing}+\epsilon^{1/2} q \eta(t)= \omega_0\hbar\left(\tilde{H}_{\ing}(N)+\frac{\epsilon^{1/2}}{m^{\frac{2-3\gamma}{4}}\alpha^{\frac{3}{4}(2-\gamma)}\hbar^{\frac{3\gamma}{2}-1}} \tilde{q}\eta(t)\right)
\end{equation}
Finally, we rescale time $t\rightarrow \omega_0^{-1}t$, such that $\eta(t)\rightarrow \omega^{1/2}_0\eta(t)$, to find
\begin{equation}
H_{\ing}+\epsilon^{1/2} q \eta(t)= \omega_0\hbar\left(\tilde{H}_0(N)+\left(\frac{\epsilon}{m^{1-\gamma}\alpha^{2-\gamma}\hbar^{2\gamma-1}}\right)^{1/2}\tilde{q}\eta(t)\right),
\end{equation}
where $\tilde{q}=\tilde{q}(N,e^{i\Theta})$ is a a-dimensional operator. Let us check the dimensions of the factor that normalizes $\epsilon$: by writing the dimension of $m$ as $\text{energy}\cdot \text{time}^2\cdot \text{length}^{-2}$ and recalling that the dimension of $\alpha$ is $\text{energy}\cdot \text{length}^{-\nu}$, we find that the factor scales as it should be 
$$
(\text{energy}\cdot \text{time}^2\cdot \text{length}^{-2})^{1-\gamma}(\text{energy}\cdot \text{length}^{\frac{2\gamma}{\gamma-2}})^{2-\gamma}
(\text{energy}\cdot \text{time})^{2\gamma-1}=\frac{\text{energy}^2\cdot \text{time}}{\text{length}^2}.
$$

In the Bohr-Sommerfeld quantization we can find how the perturbation part $\tilde{q}(N,e^{i\Theta})$ depends on $N$. From the derivation of the explicit action variable, given in Eq.~\eqref{eq:HclI}, we know that
\begin{equation}
q\propto (H_{\cl,\ing}/\alpha)^{1/\nu}\propto (\alpha m)^{\frac{\gamma-2}{4}} I^{\frac{2-\gamma}{2}} \equiv v(m,\alpha,\nu) I^{\mu},
\label{eq:vfun}
\end{equation}
with $\mu=\frac{2}{2+\nu}=\frac{2-\gamma}{2}$.
Therefore, in the Bohr-Sommerfeld quantization, $I=\hbar N$, we have
\begin{equation}
\tilde{q}(N,e^{i\Theta})\propto N^{\mu}.
\end{equation}

\section{Fokker Planck equation for and time-averaging}

This appendix contains some detailed calculations which were used in the derivation of Eq.~\eqref{eq:EigPro0}. In order to avoid confusion, we use the following summation law: all indices with enumerated subscript ($n_1,n_2,n_3\dots$, except of $n_0$  which is defined in the text) are summed over, whereas all the others ($n$, $n'$, $m$, etc.) are free. 

\subsection{Fokker Planck equation}
\label{sec:FP}

The derivation of a Fokker Planck equation from a Langevin equation is a standard procedure. We find in the Stratonovitch convention

\begin{multline}
\frac{\partial P}{\partial t}=\left\{-\mathcal{L}_{n_1n_2n_3n_4} C^{N}_{n_4n_3}\frac{\partial}{\partial C^{\Theta}_{n_1n_2}}
-i\mathcal{J}_{n_1n_2n_3n_4} \frac{\partial}{\partial C^{\Theta}_{n_1n_2}} C^{\Theta}_{n_4n_3} \right.\\ \left. 
+\frac{\tilde{\epsilon}}{2}\mathcal{F}_{m_1m_2m_3m_4}(t)\mathcal{F}_{n_1n_2n_3n_4}(t)C^{\Theta}_{n_4n_3}C^{\Theta}_{m_4m_3}\frac{\partial^2}{\partial C^N_{n_1n_2}\partial C^N_{m_1m_2}} \right\}P.
\label{eq:FP0}
\end{multline}
Since the equation is homogeneous in $C_{nn'}$, we can multiply it by $C_{nn'}C_{mm'}$ and take average over the noise to find a closed set of equations:  
\begin{equation}
\frac{d}{dt}\langle C_{n n'}^{N} C_{m m'}^{N}\rangle =  \tilde{\epsilon}\mathcal{F}_{n n' n_3 n_4}(t) \mathcal{F}_{m m' m_3 m_4}(t) \langle   C_{n_4 n_3}^{\Theta} C_{m_4m_3}^{\Theta}\rangle,
\end{equation}
\begin{multline}
\frac{d}{dt}\langle C_{nn'}^{\Theta} C_{mm'}^{\Theta} \rangle = \mathcal{L}_{n n' n_3 n_4 } \langle  C_{n_4 n_3}^{N}C_{mm'}^{\Theta}\rangle
+\mathcal{L}_{m m' m_3m_4}\langle C_{nn'}^{\Theta}  C^{N}_{m_4 m_3} \rangle \\
+i\mathcal{J}_{n n' n_3 n_4 } \langle  C_{n_4 n_3}^{\Theta}C_{mm'}^{\Theta}\rangle
+i\mathcal{J}_{m m' m_3m_4}\langle C_{nn'}^{\Theta}  C^{\Theta}_{m_4 m_3} \rangle,
\end{multline}
\begin{equation}
\frac{d}{dt} \langle  C_{n n'}^N C_{m m'}^{\Theta} \rangle  = \mathcal{L}_{m m' m_3m_4} \langle C_{n n'}^{N}  C^{N}_{m_4 m_3} \rangle
+i\mathcal{J}_{m m' m_3m_4} \langle C_{n n'}^{N}  C^{\Theta}_{m_4 m_3} \rangle.
\end{equation}

\subsection{ Magnus expansion}
\label{app:Magnus}

The growth of vectors and operators in the classical and quantum tangent spaces are governed by linear relations--- Eq.~\eqref{eq:setC} and Eqs.~\eqref{eq:ODE1}-\eqref{eq:ODE4} respectively. These equations are of the form $\dot{x^2}=M(t)x^2$. We might relax the time-dependency of $M(t)$ by employing time-averaging if the resulting growth rate, i.e., the Lyapunov exponent, is much smaller than typical rate of $M(t)$, i.e., the frequency of motion around the torus in the classical case. Technically, the elimination of high-frequency terms is made in a systematic way with the Magnus expansion, of which we need here only the first order correction. Quantum mechanically this is also possible, as we show now.  The main conclusion is the following:  time-averaging approximation is valid when $\tilde{\lambda}_{Q} \ll \omega_{n_0}$, and results in neglecting all the oscillating terms of $\mathcal{F}(t)\mathcal{F}(t)$ in Eq.~\eqref{eq:ODE1}.

Formally, Eqs.~\eqref{eq:ODE1}-\eqref{eq:ODE4} can be written as $\dot{\vec{C}}=\vec{\mathcal{S}}(t)\odot\vec{C}$, where $\vec{\mathcal{C}}$ and $\vec{\mathcal{S}}$ are respectively vector and matrix of superoperators. A formal solution to this equation is 
$$
\vec{\mathcal{C}}(t)=\mathcal{T} \left\{ e^{\int^t_0 \vec{\mathcal{S}}(t')\odot dt'}\right\}\vec{C}(0).
$$
The superoperators oscillate over time through the quantum unitary evolution of $e^{i\Theta}$, given by the unperturbed Hamiltonian $U^{\dagger}e^{i\Theta}U$, with $U\equiv e^{ i\frac{\omega_0\hbar \tilde{H}_{\ing}(N)}{\hbar}\omega_0^{-1}t}$. In principle, whenever the term $e^{im\Theta}$ appears it gives rise to 
$$
e^{im\Theta}_{nn'}=
e^{i(E_n-E_{n-m})t}\delta_{n,n'+m}=e^{i\sum^{m-1}_{r=0}\omega_{n-r}t}\delta_{n,n'+m}
$$
when evaluated in the unperturbed eigenbasis of $N$. 

For simplicity, let us assume that there is only one frequency $\omega$, $\vec{\mathcal{S}}(t)=\vec{\mathcal{S}}(\omega t)$.
Using the Magnus expansion, the solution up to some finite time $m T$, with the period $T=2\pi/\omega$, is given by
$$
\Pi^m_{i=1} e^{\vec{\mathcal{S}}_{\rm av}\odot},
$$
where the  averaged propagator is
\begin{equation}
e^{\vec{\mathcal{S}}_{\rm av}\odot}=1+\int^{T}_0 dt_1 \vec{\mathcal{S}}(\omega t_1)\odot+
\frac{1}{2}\int_0^{T}dt_1\int_0^{t_1}dt_2\left[ \vec{\mathcal{S}}(\omega t_1)\odot,\vec{\mathcal{S}}(\omega t_2)\odot\right]+\cdots.
\end{equation}
If we rescale the time in the integral with $\omega$, then the outer integral runs from 0 to $2\pi$ and the $n-$th term gives a factor of $\omega^{-n}$. Therefore, for $\omega \gg \lambda_{Q}$ we can approximate $e^{\vec{\mathcal{S}}_{\rm av}\odot}\approx 1+T \overline{\vec{\mathcal{S}}(t)\odot}$, and the corresponding general solution
$$
\vec{C}(t)= e^{\overline{\vec{\mathcal{S}}(t)\odot}} \vec{C}(0),
$$
where the overline indicates taking only the non-oscillating terms of the operation.

\bibliography{merged}

\end{document}